\documentclass{article}
\usepackage{graphicx}	 
\usepackage{amsfonts} 	
\usepackage{amsmath} 	
\usepackage{amsthm}	
\usepackage{textcomp} 

\newtheorem{defn}{Definition}
\newtheorem{ex}{Example}
\newtheorem{lem}{Lemma}
\newtheorem{thm}{Theorem}
\newtheorem{cor}{Corollary}
\begin{document}

\title{A Separation theorem for Chain Event Graphs}

\author{Peter Thwaites$^1$ and Jim Q. Smith$^2$
	\\ ${}^1$University of Leeds \hskip3mm ${}^2$University of Warwick }
\date{}
\maketitle

\begin{quotation}
\noindent{{\bf Abstract: }{\rm Bayesian Networks (BNs) are popular graphical models for the representation of statistical problems embodying
dependence relationships between a number of variables. Much of this popularity is due to the d-separation theorem of Pearl and Lauritzen, which allows
an analyst to identify the conditional independence statements that a model of the problem embodies using only the topology of the graph. However
for many problems the complete model dependence structure cannot be depicted by a BN. The Chain Event Graph (CEG) was introduced for these types
of problem. In this paper we introduce a separation theorem for CEGs, analogous to the d-separation theorem for BNs, which likewise allows an
analyst to identify the conditional independence structure of their model from the topology of the graph.}}
\end{quotation}

\begin{quotation}
\end{quotation}

\section{Introduction}

If the DAG (directed acyclic graph) of a Bayesian Network (BN) has a vertex set $\{X_1, X_2, \ldots , X_n \}$,
then there are $n$ conditional independence assertions which can simply be read off the graph. These are the
properties that state that a vertex-variable is independent of its non-descendants given its parents (the 
directed local Markov property~\cite{lauritzenbook}). Answering most conditional independence queries however,
is not so straightforward.
The d-separation theorem for BNs was first proved by Verma and Pearl~\cite{VandP}, and an alternative
version considered in \cite{Laurdsep,lauritzenbook,Cowelletal}. The theorem addresses whether the
conditional independence query $A\amalg B\ |\ C\ ?$ can be answered from the topology of the 
DAG of a BN, where $A, B, C$ are disjoint subsets of the set of vertex-variables of the DAG.
Separation theorems have been proved for more general classes of graphical model including
chain graphs~\cite{BandMilan}, alternative chain graphs~\cite{AMP2}, and ancestral graphs~\cite{TomandSp}.

However, for many problems the available quantitative dependence information cannot all be embodied in the
DAG of a BN. The Chain Event Graph (CEG) was introduced in 2006~\cite{ipmu,pgm} for the representation \& analysis of precisely these 
sorts of problems. There have been a dozen papers on CEGs published since then, principally concerned with their use for problem
representation (eg.~\cite{PaulandJim}), probability propagation~\cite{uai2008}, learning
and model selection~\cite{Guy,Guy2,Lorna2,SilanderCEG,Lorna1,RobandJim}, and causal analysis~\cite{CausalAI,newcap}.
The motivation for the development of this class is that CEGs are probably the most natural graphical
models for discrete processes when elicitation involves questions about how situations might unfold. Although the
topology of these graphs is more complicated than that of the BN, they are more expressive, as they allow us
to represent all structural quantitative information within the graph itself. Context-specific symmetries which are 
not intrinsic to the structure of the BN \cite{kollerUAI1996,McAllester,PooleandZ,SalCanM} are fully expressed in the 
topology of the CEG, which also recognises logical or structural zeros in probability tables, and the numbers of levels taken
by problem-variables. This last has been found to be essential to understanding
the geometry of BN models with hidden variables~\cite{Allmanrhodes,MondandJim}. In this paper we return to the mathematics underpinning CEG models,
and provide a separation theorem for these graphs.

The CEG is a tree-based graphical structure with a passing resemblance to graphs such as Bozga \& Malers' probabilistic
decision graph~\cite{FirstPDGs} (made popular by Jaeger et al in~\cite{Jaeger04}). It differs from these in that edges in a CEG label events that 
might happen to an individual in a population given
a particular partial history, and the coalescing of vertices \& colouring of edges together encode conditional independence/Markov structure. The colouring
of CEGs and their acyclicity also distinguish them from Markov state space diagrams. Finite CEGs as discussed in this paper also have finite 
event spaces whose atoms correspond to the distinct possible histories or developments that individuals in a population might have. The tree-structure
imparts to these atoms an additional longitudinal element consisting of the stages of an individual's development. 
We note in passing that colour has recently been found to provide a valuable embellishment to other graphical
models (see for example~\cite{HandL}). 

Even more so than is the case with BNs, there are a number of conditional independence
properties which can simply be {\it read} off the CEG~\cite{PaulandJim}, and given the
tree-based nature of the CEG these properties are naturally context-specific. That is to say they are properties of the form
$A \amalg B\ |\ \Lambda$ for some event $\Lambda$. An example would be that a particular lifestyle-related medical condition is independent of gender given
that the subject is a smoker.
An analogous statement for a discrete BN would be of the form
$$p(A \ |\ B, C = \boldsymbol{c}) = p(A \ |\ C = \boldsymbol{c})$$ 
for some subsets of variables $A, B, C$, some {\bf specific} vector
value $\boldsymbol{c}$ of $C$ and all vector values of $A$ and $B$. 
The class of conditioning
events we can tackle with a CEG is however much richer than that generally considered when using BN-based analysis.

In Section 2 we use a toy example to introduce CEGs. A naive criticism of tree-based graphical structures is that they will be too complex for
larger problems. We note that the {\it picture} is simply for the investigator's (or a client's) benefit: as with any large system, analysts need to 
consider both local and global aspects -- the full CEG for a large
problem may exist only as a set of computer constraints; local aspects of the problem can be drawn out as a simple graph. Our example here is small so that
we can use it effectively to illustrate key ideas.

We then formally define a CEG, and explain how coalescence \& colouring encode conditional independence structure. Events \& random variables defined on 
CEGs are introduced through our example, as are sub-CEGs conditioned on an event of interest.

In Section 3 we introduce elementary variables associated with the vertices of the CEG, and use these to construct a separation theorem. Comparison
with the d-separation theorem for BNs is made through corollaries and our running example. Section~4 develops some of the ideas from earlier sections.

\section{Chain Event Graphs}	

Definitions of Chain Event Graphs (of varying degrees of complexity) have appeared in many of the previous papers on these graphs.
We offer a detailed formal definition here so that the theorems in later sections have a firm mathematical foundation.

\subsection{Event Trees}	

We introduce CEGs in this section through the use of a toy example, simple enough to illustrate the key ideas.

\smallskip
The CEG is a function of an {\it event tree}~\cite{Shafer}, and was created to overcome some of the shortcomings of these graphs. So we start by 
considering an event tree elicited from some expert.

\begin{ex}			
A researcher is investigating a population of people whose parents sufferered from an inherited medical condition {\rm C}. She has information on
the gender of each individual; if and when they displayed a symptom {\rm S} (never, before puberty, after puberty); and whether or not they developed
the condition~{\rm C}. Her current research is retrospective so she also has these individual's ages at death. She suspects 
that the condition can lead to early death, so she produces an indicator that for each individual
records whether or not they died before the age of 50.
She knows that an individual who does not display symptom {\rm S} will not develop the condition.

An event tree for this information is given in Figure~1.
\end{ex}

\begin{figure}[t]	
\includegraphics[width=12cm,height=9cm]{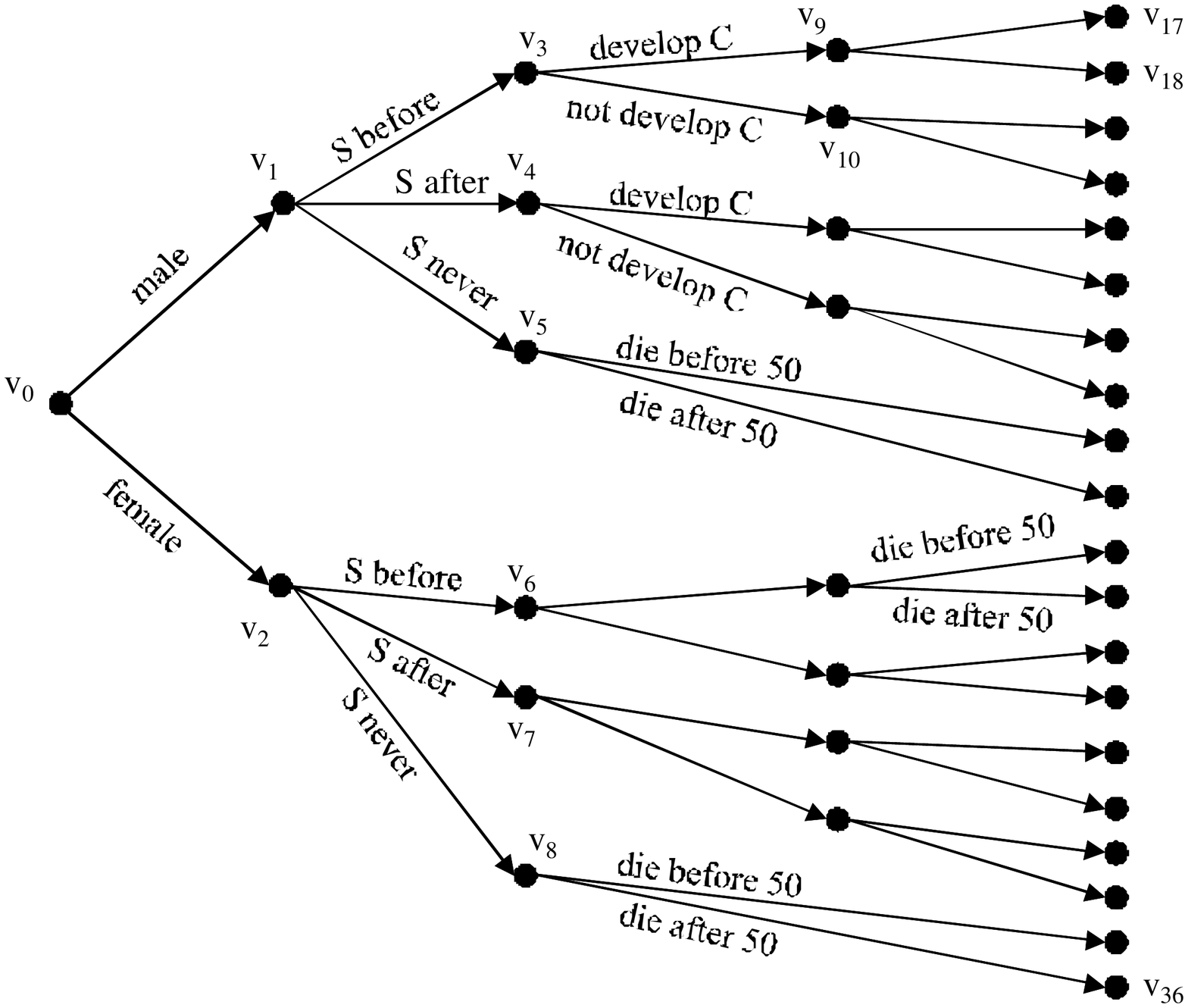}
\caption[]{Event tree for Example 1}
\end{figure}

The tree is a natural description of the problem. The labels on the edges of each root-to-leaf path (eg. {\it male, display {\rm S} before puberty,
develop {\rm C}, die before~50}) in Figure~1 follow a temporal order, and the absence of {\it condition} edges in the 9th, 10th, 19th \& 20th such paths reflects
the expert's knowledge that individuals who do not display S do not develop C.

\smallskip
An event tree $\cal{T}$ is a connected directed graph with no cycles. It has one root vertex ($v_0$) with no parents, whilst all other vertices have 
exactly one parent. A leaf-vertex is a vertex with no children. We denote the vertex set of $\cal{T}$ by~$V(\cal{T})$ and the edge set by
$E(\cal{T})$. A directed root-to-leaf path in $\cal{T}$ is called a {\it route}.

Although an event tree could be used to represent an observer's beliefs about the possible developments of some individual, we make the
assumption that the tree relates to a population. Hence the routes of the tree describe precisely the possible 
developments or histories that an individual in the population can experience. This description takes the form of a sequence of edge-labels, each 
describing what can happen next at a vertex. So in Figure~1 for example, an individual who is male reaches vertex $v_1$ where the possible
immediate developments are that he displays S before puberty, after puberty or not at all.

We specify that the edges leaving any vertex in the tree must have distinct labels; that each individual can only pass along one edge leaving any
vertex, and the choice of which edge is determined only by the variable controlling the next stage of development (eg. {\it symptom}) and not by
any possible further developments downstream of these edges (ie. towards the leaves).

We also require that each route corresponds to a real possible development or history of an individual in the population. So each such path has
a non-zero probability that some individual might take this path. Also, the number of routes corresponds exactly to the number of distinct possible
histories or developments (defined by the edge-labels) that some individual could experience.

Once we have a set of routes, and some ordering on these paths, then the edge-labels {\bf define} the tree structure. In our
example the first variable in our order is {\it gender}, so $v_0$ has two emanating edges, labelled {\it male} \& {\it female}. The second variable 
is {\it symptom}, so $v_1$ \& $v_2$ both have three emanating edges, labelled {\it displays {\rm S} before puberty, displays {\rm S} after puberty}
\& {\it never displays} S. We know that individuals who never display S will not develop condition C, so the edges emanating from $v_5$ \& $v_8$
label the possible values of the life-expectancy indicator, whereas those emanating from $v_3,\ v_4,\ v_6$ \& $v_7$ label the possible values
of {\it condition}.

In this paper we use the notation $\lambda$ to denote a route, and the set of routes of $\cal{T}$ is labelled $\Lambda(\cal{T})$. When the tree
is applied to a population, each route $\lambda$ corresponds to a possible history or development of an individual in the population, and
hence to an atom in an event space defined by the tree. The sigma field of events associated with $\cal{T}$ is then the set of all possible
unions of atoms $\lambda$ in $\Lambda(\cal{T})$. Note that the tree encodes an additional longitudinal development or history for the individual,
not encoded by the sigma field alone~\cite{Shafer}. Events in the sigma field of the tree are denoted $\Lambda$.

\smallskip
So for instance, in Example~1 the event $\Lambda$ corresponding to {\it displayed {\rm S} before puberty and died 
before the age of~50} is simply the union of the 1st, 3rd, 11th \& 13th routes in the tree in Figure~1.

\bigskip
\noindent{{\bf Example 1 continued.} {\it Our researcher has done sufficient analysis of the data to tell us that:}}

$\bullet$ {\it life expectancy of individuals in this population is independent of gender given that S is not displayed,}

$\bullet$ {\it males who display S at any point and females who display S before puberty have the same joint probability distribution over the variables 
{\rm condition} and {\rm life expectancy}.}

\noindent{\it Moreover}

$\bullet$ {\it if and when an individual displays {\rm S} is independent of gender,}

\noindent{\it and she believes that}

$\bullet$ {\it males and females who display {\rm S} at any point have the same probability of developing the condition.}

\smallskip
It is the fact that traditional trees cannot readily depict this sort of information which has led to tree-based analysis not receiving the attention 
it deserves. It is actually relatively easy to portray these types of conditional independence or Markov properties on a tree -- all we need to do
is add colour to the edges, as in Figure~2 (where edges with the same colouring carry the same probability).

\begin{figure}[t]	
\includegraphics[width=12cm,height=9cm]{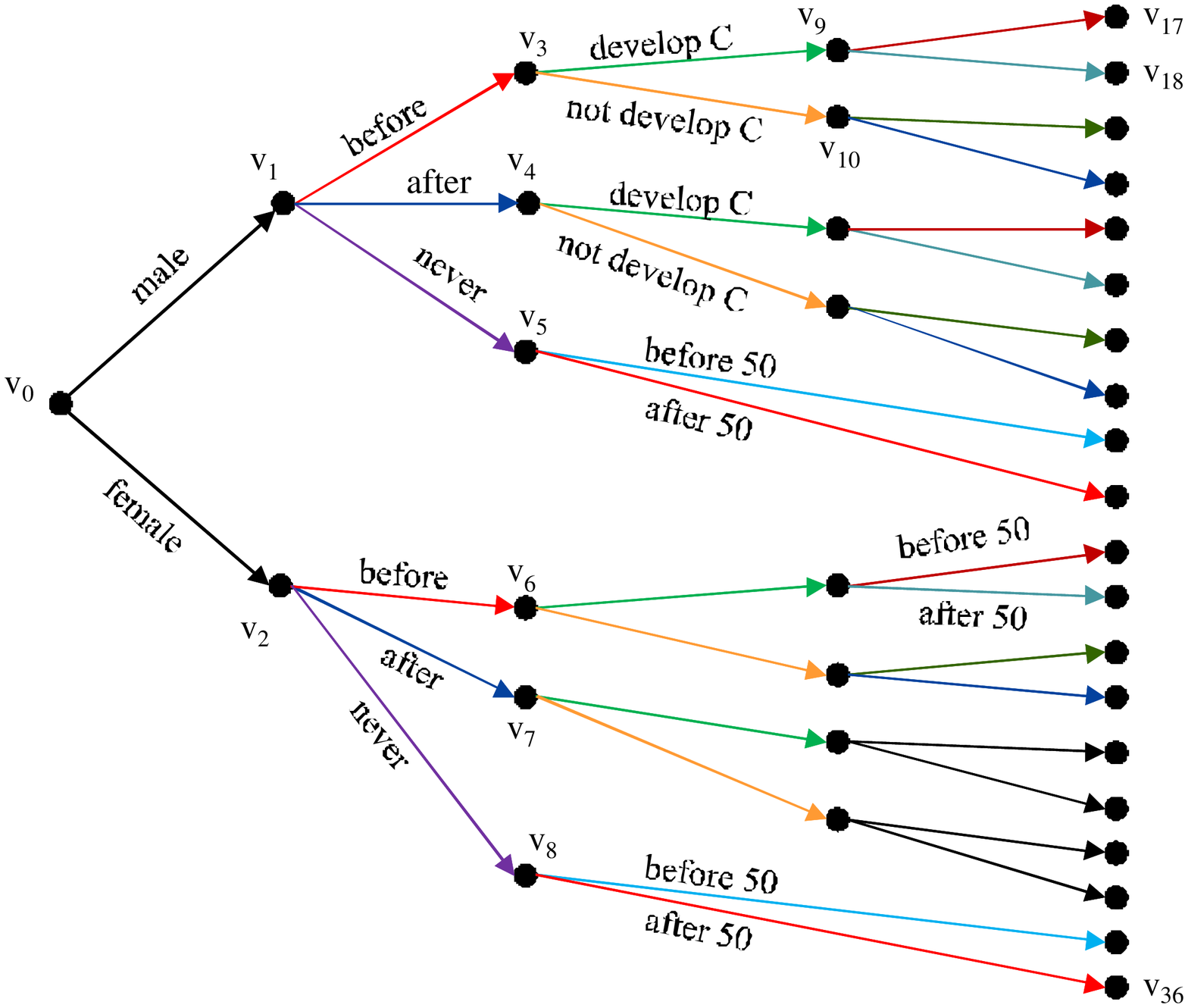}
\caption[]{Coloured event tree for Example 1}
\end{figure}

Despite the colouring this is still a rather cumbersome representation. To make it more compact we use the idea of coalesced trees, used in decision 
analysis since~\cite{Olmstead}. In a coalesced event tree vertices from which the sets of possible complete future developments have the same
probability distribution are coalesced.

\smallskip
So in the tree in Figure 2 we can coalesce the vertices $v_3, v_4$ \& $v_6$ and also the vertices
$v_5$ \& $v_8$ (the vertices $v_9, v_{11}$ \& $v_{13}$ and $v_{10}, v_{12}$ \& $v_{14}$ are also coalesced, but this coalescence is in a sense 
absorbed into that of $v_3, v_4$ \& $v_6$).

\begin{figure}[t]	
\includegraphics[width=12cm,height=8.2cm,bbllx=54,bblly=100,bburx=777,bbury=597]{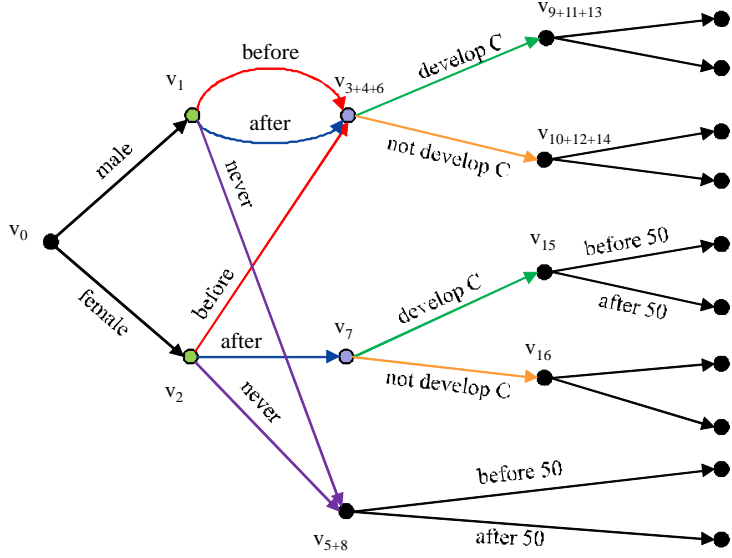}
\caption[]{Coalesced event tree for Example 1}
\end{figure}

\smallskip
The combination of colouring and coalescence gives us a more compact graph that allows us to portray all conditional independence properties
of the type described in Example~1 above.

\smallskip
So in Figure 3, the first two of the four statements provided by our researcher are depicted by the coalescence, but the latter two require
the colouring of the edges leaving $v_1$ \& $v_2$ and $v_{3+4+6}$ \&~$v_7$. 
The colouring of the edges emanating from $v_{9+11+13}$ \& from $v_{10+12+14}$ is suppressed as it no longer yields any extra information.

\subsection{Probabilities on Trees}	

In section 2.1 we talked in general terms about probabilities on trees. In this section we formalise these ideas.

\bigskip
\noindent{In Figure~1 the probability of the atom \{{\it male, displayed S before puberty, developed C, died before~50}\}
is clearly:}
\begin{align*}
p(\hbox{male}) &\times p(\hbox{displayed S before puberty}\ |\ \hbox{male}) \\
	&\times p(\hbox{developed C}\ |\ \hbox{male, displayed S before puberty}) \\
	&\times p(\hbox{died before}\ 50\ |\ \hbox{male, displayed S before puberty, developed C}) \\
	\intertext{which can be written as}
&\pi_e(v_1\ |\ v_0)\ \pi_e(v_3\ |\ v_1)\ \pi_e(v_9\ |\ v_3)\ \pi_e(v_{17}\ |\ v_9)\ 
\end{align*}  
\noindent{where $\pi_e(v_{17}\ |\ v_9)$ is the probability of an individual having reached the vertex~$v_9$ in Figure~1 (ie. they are male,
displayed~S before puberty \& developed~C) then taking the edge $e(v_9, v_{17})$ to reach the vertex $v_{17}$ (ie. they die before~50) etc.}

\bigskip
We can assign a probability to each atom of the event space as:
$$p(\lambda) = \prod_{e(v, v') \in \lambda} \pi_e(v'\ |\ v)$$
where $e(v, v')$ means the edge from vertex~$v$ to vertex~$v'$, $e(v, v') \in \lambda$ means that $e(v, v')$ lies on the route $\lambda$, and
$\pi_e(v'\ |\ v)$ is the conditional probability of traversing the edge $e(v, v')$ given that have reached the vertex $v$.

We call the probabilities $\{ \pi_e(v'\ |\ v)\}$ the {\it primitive} probabilities of the tree~$\cal{T}$.

The set $\{ p(\lambda)\}$ defines a probability measure over the sigma field of events formed by the atoms $\lambda \in \Lambda(\cal{T})$.
Strictly speaking these probabilities are the fundamental probabilities of the system as they are the probabilities of the atoms. Each primitive
probability has then a unique value determined by these $\{p(\lambda)\}$. The conditional probability 
of an edge $e(v, v')$ is given by:
$$\pi_e(v'\ |\ v) =\frac{\sum_{\lambda: e(v,v') \in \lambda} p(\lambda)}{\sum_{\lambda: v \in \lambda} p(\lambda)}$$
where the numerator is the sum of the probabilities of all routes utilising the edge $e(v, v')$, and the denominator is the sum of the probabilities
of all routes passing through the {\it start-vertex} $\ v$ of the edge $e(v, v')$. So for example in Figure~1 we have:
$$\pi_e(v_{17}\ |\ v_9) = \frac{p(\lambda_1)}{p(\lambda_1) + p(\lambda_2)}$$
where $\lambda_1$ is the atom corresponding to the route $v_0 \longrightarrow v_{17}$ ($\lambda(v_0, v_{17})$) and
$\lambda_2$ is the atom corresponding to the route $v_0 \longrightarrow v_{18}$ ($\lambda(v_0, v_{18})$).

In practice however, our elicitation of the tree is likely to yield primitive probabilities of the sort described above, rather than probabilities
of atoms.

\subsection{Positions and Stages}	

To allow our event tree to encode the full conditional independence structure of the model we introduce two partitions of the tree's vertices.

Let $V^0(\cal{T})$ ($\subset V(\cal{T})$) be the set of non-leaf vertices of $\cal{T}$ (called {\it situations} in~\cite{CausalAI}). 
Also let $v \prec v'$ denote that the vertex $v$ precedes the vertex $v'$ on some route.

Then for any non-leaf vertex $v_a \in V^0(\cal{T})$ and leaf-vertex $v_a'' \in V({\cal{T}})\ \backslash\ V^0(\cal{T})$ such that $v_a \prec v_a''$,
there is a unique subpath $\mu (v_a, v_a'')$ comprising of the edges of the route $\lambda(v_0, v_a'')$ which lie between the vertices $v_a$ 
and $v_a''$. 

Let:
$$\pi_{\mu}(v_a''\ |\ v_a) = \prod_{e(v, v') \in \mu(v_a, v_a'')} \pi_e(v'\ |\ v)$$

Now each vertex $v \in V^0(\cal{T})$ labels a random variable $J(v)$ whose state space ${\mathbb{J}} (v)$ can be identified with the set of 
$v \longrightarrow \hbox{\it leaf}$ subpaths $\{ \mu(v, v'') \}$.

\begin{defn}	
{\bf Positions.} For an Event Tree ${\cal{T}}(V({\cal{T}}),E({\cal{T}}))$, the set $V^0(\cal{T})$ is partitioned into equivalence classes, called 
{\rm positions} as follows:\hfill\break
Vertices $v_a, v_b \in V^0(\cal{T})$ are members of the same equivalence class {\rm (position)} if there is a bijection $\phi$ between
${\mathbb{J}} (v_a)$ and ${\mathbb{J}} (v_b)$ such that if\break $\phi: \mu(v_a, v_a'') \mapsto \mu(v_b, v_b'')$, then\hfill\break
(a) the ordered sequence of edge-labels is identical for $\mu(v_a, v_a'')$ and for $\mu(v_b, v_b'')$,\hfill\break
(b) $\pi_{\mu}(v_a''\ |\ v_a) = \pi_{\mu}(v_b''\ |\ v_b)$.
\end{defn}

Now, from section 2.1, tree structure is defined by the edge-labels, so (a) above means that the subtrees rooted in $v_a$ and $v_b$ have
identical topological structure.

Similarly, from section 2.2, our edge probabilities are uniquely defined by the route probabilities. We can see that the edge probabilities in the
subtrees rooted in $v_a$ and $v_b$ must be uniquely defined by the sets of probabilities $\{\pi_{\mu}(v_a''\ |\ v_a)\}$ and 
$\{\pi_{\mu}(v_b''\ |\ v_b)\}$. So (b) above means that the corresponding edge probabilities in these two subtrees are equal.

So two vertices in a tree are in the same {\it position} if the sets of possible complete future developments from these vertices have the same
probability distribution. We denote the set of positions of $\cal{T}$ by $P(\cal{T})$

We noted earlier that knowing this partition of vertices is insufficient for us to fully describe the conditional independence structure of
the tree, so we introduce a second partition.

Each vertex $v \in V^0(\cal{T})$ also labels a random variable $K(v)$ whose state space ${\mathbb{K}} (v)$ can be identified with the set of 
directed edges $e(v, v')$ emanating from~$v$.

\begin{defn}	
{\bf Stages.} For an Event Tree ${\cal{T}}(V({\cal{T}}),E({\cal{T}}))$, the set $V^0(\cal{T})$ is partitioned into equivalence classes, called 
{\rm stages} as follows:\hfill\break
Vertices $v_a, v_b \in V^0(\cal{T})$ are members of the same equivalence class {\rm (stage)} if there is a bijection $\psi$ between
${\mathbb{K}} (v_a)$ and ${\mathbb{K}} (v_b)$ such that if\break $\psi: e(v_a, v_a') \mapsto e(v_b, v_b')$, then $\pi_e(v_a'\ |\ v_a) = 
\pi_e(v_b'\ |\ v_b)$.
\end{defn}

So two vertices in a tree are in the same {\it stage} if their sets of emanating edges have the same probability distribution.

Note that the set of stages is coarser than the set of positions, and that vertices in the same position are necessarily in the same stage.

We also add colouring to trees to illustrate the stage structure. So vertices in the same stage are given the same colour, and edges emanating from 
vertices in the same stage are coloured according to their probabilities / labels. This induces a partition on $E({\cal{T}})$.

\subsection{Chain Event Graphs}		

The {\em Chain Event Graph} $\cal{C}$ is a directed acyclic graph (DAG), which is connected, having a unique {\em root vertex} (with no incoming 
edges) and a unique {\em sink vertex} (with no outgoing edges). Unlike the BN more than one edge can exist between two vertices of a CEG. 
The CEG also generally has its vertices and edges coloured, although most of this paper will deal with uncoloured versions called {\it Simple}
CEGs. The root and sink vertices of a CEG are labelled $w_0$ and~$w_{\infty}$.

\begin{defn}	
{\bf Chain Event Graph.} The CEG ${\cal{C}}(\cal{T})$ (a function of the tree ${\cal{T}}(V({\cal{T}}),E({\cal{T}}))$) is the graph with
vertex set $V({\cal{C}})$ and edge set $E({\cal{C}})$ defined by:
\begin{enumerate}
\item $V({\cal{C}}) \equiv P({\cal{T}}) \cup \{ w_{\infty} \}$;
\item \begin{enumerate} \item For $w, w' \in V({\cal{C}}) \backslash \{ w_{\infty} \}$ there is a directed edge $e(w, w') \in E({\cal{C}})$ iff there 
	are vertices
	$v, v' \in V^0({\cal{T}})$ such that the vertex $v$ is in the position $w$ ($\in P({\cal{T}})$), $v'$ is in the position $w'$ 
	($\in P({\cal{T}})$), and there is an edge $e(v, v') \in  E({\cal{T}})$;
 	\item For $w \in V({\cal{C}}) \backslash \{ w_{\infty} \}$ there is a directed edge $e(w, w_{\infty}) \in E({\cal{C}})$ iff there is a vertex
	$v \in V^0({\cal{T}})$ such that $v$ is in the position $w$\break ($\in P({\cal{T}})$),  
	and there is an edge $e(v, v') \in  E({\cal{T}})$ for some leaf-vertex $v' \in V({\cal{T}}) \backslash V^0({\cal{T}})$.
\end{enumerate} 
\end{enumerate}
\end{defn}

Note that the vertex set of ${\cal{C}}(\cal{T})$ consists of the {\it positions} of $\cal{T}$ and the sink-vertex $w_{\infty}$. Positions in 
${\cal{C}}(\cal{T})$ are said to be in the same {\it stage} if the component vertices (in $\cal{T}$) of these positions are in the same stage.
Colouring in ${\cal{C}}(\cal{T})$ is inherited from $\cal{T}$. The constraints associated with the positions \& stages of a CEG hold for the entire
population to which the CEG has been applied,

\bigskip
\noindent{{\bf Example 1 continued.} {\it To convert the coalesced tree from Figure~3 to a CEG is straightforward. We simply combine the
leaf-vertices into a sink-vertex $w_{\infty}$ as in Figure~4.

\begin{figure}[t]	
\includegraphics[width=12cm,height=8.2cm,bbllx=54,bblly=100,bburx=777,bbury=597]{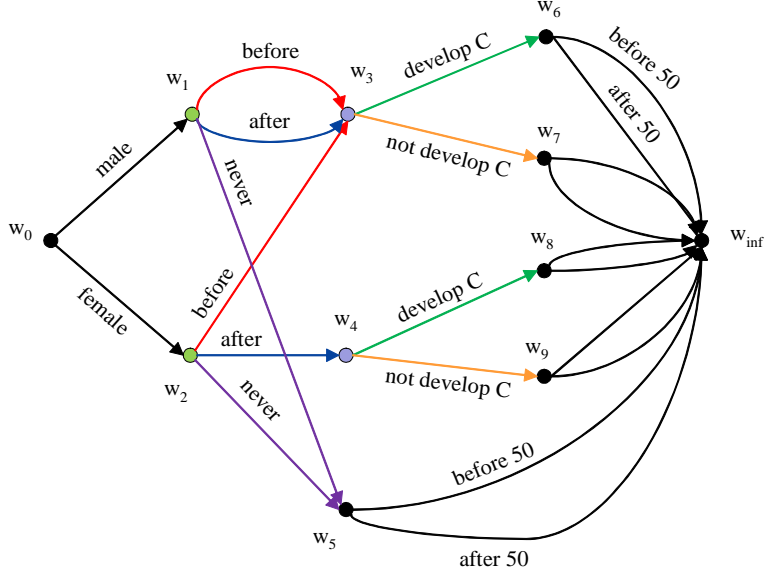}
\caption[]{CEG for Example 1}
\end{figure}

The positions here are $w_0$ through $w_9$. We note that $w_1$ \& $w_2$ are in the same stage (as can be seen from the colouring), $w_3$ \& $w_4$ 
are in the same stage, and each of $w_5$ through $w_9$ is in a stage by itself.

The position $w_3$ encodes the conditional independence / Markov property that  
males who display {\rm S} at any point and females who display {\rm S} before puberty have the same joint probability distribution over the variables 
{\rm condition} and {\rm life expectancy}. 
The position $w_5$ encodes the property that
life expectancy of individuals in this population is independent of gender given that {\rm S} is not displayed.
The stage $\{ w_1, w_2 \}$ encodes the property that
if and when an individual displays {\rm S} is independent of gender.
The stage $\{ w_3, w_4 \}$ encodes the property that {\rm condition} is independent of gender given {\rm S} displayed.}

\bigskip
Without ambiguity we simplify our notation ${\cal{C}}(\cal{T})$ to $\cal{C}$.

Analogously with the tree, a directed $w_0 \rightarrow w_{\infty}$ path $\lambda$ in $\cal{C}$ is called a {\em route}.  The set of routes 
of $\cal{C}$ is labelled $\Lambda ({\cal{C}})$.
We write $w \prec w'$ when the position~$w$ precedes the position~$w'$ on a route.

When the CEG is applied to a population, each route $\lambda$ corresponds to a possible history or development of an individual in the population, and
hence to an atom in the event space defined by the CEG. The sigma field of events associated with $\cal{C}$ is then the set of all possible
unions of atoms $\lambda$ in $\Lambda(\cal{C})$. Like the tree, the CEG encodes an additional longitudinal development or history for the individual,
not encoded by the sigma field alone. Events in the sigma field of the CEG are denoted $\Lambda$.

Note that the number of routes in the CEG equals the number in the tree and corresponds exactly to the number of possible distinct histories or
developments that some individual in the population could experience. And since no route has a zero probability, all edges in the CEG have non-zero
conditional probabilities associated with them.

Because the CEG's atoms have this implicit longitudinal development associated with them, certain events in the sigma field are particularly
important. Let $\Lambda (w)$ denote the event that an individual unit takes a route that passes through the position $w\in V(\mathcal{C})$.
$\Lambda (w, w')$ is then the union of all routes passing through the positions $w$ and $w'$, $\Lambda (e(w, w'))$ is the union of all routes 
passing through the edge $e(w, w')$, and $\Lambda (\mu (w, w'))$ is the union of all routes utilising the subpath $\mu (w, w')$.

\subsection{Probabilities on CEGs}   

As with trees, underlying the CEG there is a probability space which is specified by assigning probabilities to the atoms. For each position
$w \in V({\cal{C}}) \backslash \{ w_{\infty} \}$ and edge $e(w, w')$ emanating from $w$,
we denote by $\pi_e(w'\ |\ w)$ the probability of traversing the edge $e(w, w')$ conditional on having reached the position $w$. We call the
probabilities $\{\pi_e(w'\ |\ w) : e(w, w') \in E({\cal{C}}), w \in V({\cal{C}}) \backslash \{ w_{\infty} \} \}$ the {\it primitive probabilities} 
of $\cal{C}$.

Then, analogously with trees, for each atom $\lambda$:
$$p(\lambda) = \prod\limits_{e(w, w')\in \lambda} \pi_e(w'\ |\ w)$$
as both the atoms and the primitive probabilities are identical to the corresponding atoms and primitive probabilities
in the tree.

The set $\{ p(\lambda)\}$ defines a probability measure over the sigma field of events formed by the atoms $\lambda \in \Lambda({\cal{C}})$.

This assignment of probabilities implicitly demands a Markov property over the flow of the units through the graph. Thus, in the context 
of our medical example, the probablility of an individual with attributes ({\it male, displayed symptom $S$ before puberty}), ({\it male, displayed 
symptom $S$ after puberty}) or ({\it female, displayed symptom $S$ before puberty}) developing the condition depends only on the fact that the 
subpaths corresponding to these pairs of attributes terminate at the position $w_3$, and not on the particular subpath leading to $w_3$. The 
probability this individual develops the condition is then $\pi_e(w_6\ |\ w_3) \equiv p(\Lambda (e(w_3, w_6))\ |\ \Lambda (w_3))$. So we only 
need to know the position a unit has reached in order to predict as well as is possible what the next unfolding of its development will be.

This Markov hypothesis looks strong but in fact holds for many families of statistical model. For example all event tree descriptions of a 
problem satisfy this property, all finite state space context specific Bayesian Networks as well as many other structures~\cite{PaulandJim}.

We can go further and state that the sets of possible future developments (whether or not they developed
the condition {\bf and} whether or not they died before the age of 50) for individuals taking any of these three subpaths
must be the same. Moreover the conditional probability of any particular subsequent development must be the same
for individuals taking any of these three subpaths. 

Note also that if positions $w_a$ and $w_b$ are such that the sets of possible future developments from $w_a$ and $w_b$ are identical, 
and the conditional joint probability distributions over these sets are identical, then $w_a$ and $w_b$ are the {\bf same} position, and must be
coalesced for our graph to be a CEG.
 
The probability of any event $\Lambda$ in the sigma field is hence of the form
$$p(\Lambda) = \sum_{\lambda \in \Lambda} p(\lambda) = \sum_{\lambda \in \Lambda} \prod_{e(w, w') \in \lambda}
	\pi_e (w'\ |\ w)$$
where $\lambda \in \Lambda$ means that $\lambda$ is one of the component atoms of the event $\Lambda$.\break
In this paper we will also use the following further notation:

\noindent{$\pi_{\mu} (w'\ |\ w) \equiv p (\Lambda (\mu (w, w'))\ |\ \Lambda (w))$ denotes the probability of
utilising the subpath $\mu (w, w')$ (conditional on passing through $w$),}
 
\noindent{$\pi (w'\ |\ w) \equiv p (\Lambda (w, w')\ |\ \Lambda (w))
= \sum_{\mu} \pi_{\mu} (w'\ |\ w)$ denotes the probability of arriving at $w'$ conditional on passing through~$w$.}

\smallskip
Expressing a problem as a CEG allows domain experts to check their beliefs in a very straightforward manner:

\smallskip
We stated in Section 2.1 that the expert in our example believed that males \& females who display {\rm S} at any point
have the same probability of developing~{\rm C}. This is depicted in the colouring of the edges emanating from $w_3$ \& $w_4$ in Figure~4. Our expert
can now use the techniques developed in~{\rm\cite{Guy,Guy2,Lorna2,SilanderCEG}} to test the model represented by Figure~4 against alternative models 
with different
conditional independence / Markov structure. Such a test might yield information that grouped the vertices $v_3, v_4, v_6$ \& $v_7$ from Figure~1 into
different positions than those in Figure~4; or that the vertices $v_3, v_4$ \& $v_6$ are
indeed in the same stage and position, but that the vertex $v_7$ is not in this stage (ie. the probability of developing {\rm C} is
different for females who display {\rm S} after puberty), and so the edges leaving $w_3$ \& $w_4$ in Figure~4 would no longer have the same colouring.

\subsection{Conditioning on events} 

Most conditional independence queries that could realistically be of interest to an analyst can be answered purely by inspecting the topology 
of a CEG. And most of these queries involve conditioning on what is known as an intrinsic event.

\begin{defn}{\bf Intrinsic events.}
Let ${\mathcal{C}}_{\Lambda}$ be the subgraph of $\mathcal{C}$ consisting of only those positions and edges that lie on a route $\lambda \in
\Lambda$, and the sink-vertex $w_{\infty}$. $\Lambda$~is \emph{intrinsic to} $C$ if the number of $w_0 \longrightarrow w_{\infty}$ paths in
${\mathcal{C}}_{\Lambda}$ equals the number of atoms in the set $\{ \lambda \}_{\lambda \in \Lambda}$.
\end{defn}

The idea of intrinsic events is closely related to that of faithfulness in BNs~\cite{Meek,Spirtes}. Note that each atom (\& therefore edge) in
${\mathcal{C}}_{\Lambda}$ must by construction have a non-zero probability, but edge-probabilities in ${\mathcal{C}}_{\Lambda}$ may differ from
those in ${\mathcal{C}}$ since some vertices in ${\mathcal{C}}_{\Lambda}$ will have fewer emanating edges than they have in ${\mathcal{C}}$, and
the probabilities on the emanating edges of any vertex must sum to one. 

All atoms of the sigma field of $\cal{C}$ are intrinsic, as are $\Lambda(w)$, $\Lambda(w,w')$, $\Lambda(e(w,w'))$, $\Lambda(\mu(w,w'))$
(provided these are non-empty), and as is the exhaustive set $\Lambda (w_{0})$. If we include the empty set in the set of intrinsic events then 
we note that intrinsic sets are closed under intersection and so technically form a $\pi$-system (see for example~\cite{kallenberg}) we can 
associate with the CEG~$\mathcal{C}$. 

Not all events in the sigma field are necessarily intrinsic, because the class of intrinsic events is not closed under union. For example, for the 
CEG in Figure~4, the event $\Lambda$ consisting of the union of the two atoms described by the routes {\it male, display {\rm S} before puberty,
develop {\rm C}, die before 50} and {\it male, display {\rm S} after puberty, develop {\rm C}, die after 50} produces a 
subgraph~${\cal{C}}_{\Lambda}$ which has four distinct routes, so $\Lambda$ is not intrinsic.  However, our interest in intrinsic events is that we 
can condition on them, and we show below that conditioning on intrinsic events often destroys the stage-structure of $\cal{C}$. Conditioning on
non-intrinsic events usually destroys position-structure. From this we argue that if we know that we wish to condition on an event such as the 
one described above, we would simply sacrifice the position-structure of our CEG (knowing that it would probably be lost in the conditioning
anyway) and split (uncoalesce) the position $w_3$ to form a graph for which this event is intrinsic. 

Even without such sleight of hand, the class of intrinsic events is rich enough to encompass virtually all of the conditioning events in the
conditional independence statements we would like to query. In particular, if our model can be expressed as a BN (with vertex-variables
$\{ X_j \}$) then any set of observations expressible in the form $\{ X_j \in A_j \}$ ($\{ A_j \}$ subsets of the sample spaces of $\{ X_j \}$) 
is a proper subset of the set of intrinsic events defined on the CEG of our model~\cite{uai2008}.

\bigskip
\noindent{{\bf Example 1 continued.} {\it Suppose for illustrative convenience that the edges labelled \emph{male, female, displayed S before
puberty, displayed S after puberty, never displayed S, developed C, did not develop C} in our CEG have the probabilities $\frac{1}{2}$, $\frac{1}{2}$, 
$\frac{1}{4}$, $\frac{1}{4}$, $\frac{1}{2}$, $\frac{1}{2}$, $\frac{1}{2}$. Now let us condition on the event $\Lambda$ which is the union of
all routes {\bf except} \{{\rm female, displayed S after puberty, did not develop C, died before 50}\} and \{{\rm female, displayed S after 
puberty, did not develop C, died after 50}\}. This event $\Lambda$ is clearly intrinsic to $\cal{C}$, and has the probability 
$p(\Lambda)~=~\frac{15}{16}$.}}

{\it When we condition on $\Lambda$, the routes $\lambda$ which are components of $\Lambda$ get new probabilities $p(\lambda\ |\ \Lambda) =
p(\lambda, \Lambda) / p(\Lambda) = p(\lambda) / p(\Lambda)$. In this case each route has its probability multiplied by $\frac{16}{15}$. We
leave it as a (simple) exercise to show that all edge-probabilities remain unchanged except:}
\begin{align*}
\pi_e(w_1\ |\ w_0)\ &\hbox{becomes}\ 8/15 \\
\pi_e(w_2\ |\ w_0)\ &\hbox{becomes}\ 7/15 \\
\pi_e(w_3\ |\ w_2)\ &\hbox{becomes}\ 2/7 \\
\pi_e(w_4\ |\ w_2)\ &\hbox{becomes}\ 1/7 \\
\pi_e(w_5\ |\ w_2)\ &\hbox{becomes}\ 4/7 \\
\pi_e(w_8\ |\ w_4)\ &\hbox{becomes}\ 1 \\
\hbox{the edge}\ \pi_e(w_9\ |\ w_4)\ &\hbox{does not exist in}\ {\cal{C}}_{\Lambda}
\end{align*}
\noindent{\it So the positions $w_1$ and $w_2$ are no longer in the same stage.}

\bigskip
So, as already noted, conditioning on an intrinsic event can destroy stage-structure. This leads us to define an uncoloured version of the CEG.

\begin{defn}{\bf Simple CEG}	
A simple CEG (sCEG) is a CEG where there are no constraints on edge-probabilities, except that (i)~all edge-probabilities must be greater than zero
(a consequence of the requirement we made for trees), and
(ii)~the sum of emanating-edge-probabilities for any position must equal one.
\end{defn}

What this means in practice is that stage-structure is suppressed: there are no stages which are not positions, and so colouring is redundant.
There is an analogy here with BNs to which one can always {\bf add} edges, and sacrifice a little conditional independence structure.

We show now that the class of sCEG models is closed under conditioning on an intrinsic event:

\begin{thm}	
For an event $\Lambda$, intrinsic to $\cal{C}$, the subgraph ${\cal{C}}_{\Lambda}$ is an sCEG. If the probability of any route $\lambda$ in the
sigma field of ${\cal{C}}_{\Lambda}$ is given by $p_{\Lambda}(\lambda) = p(\lambda\ |\ \Lambda)$, then the edge-probabilities in ${\cal{C}}_{\Lambda}$ 
are given by:
$$\hat{\pi}_e (w'\ |\ w) = \frac{p (\Lambda\ |\ \Lambda (e(w, w')))}{p (\Lambda\ |\ \Lambda (w))}\ \pi_e (w'\ |\ w)$$
\end{thm}

The proof of this theorem is in the appendix. We note that this result has been successfully used to develop fast propagation algorithms for 
CEGs~\cite{uai2008}.

Note that the probability of an atom $\lambda$ in $\cal{C}$ conditioned on the intrinsic event $\Lambda$ is 
the probability of that atom in the sCEG ${\cal{C}}_{\Lambda}$ (denoted $p_{\Lambda}(\lambda)$). It is then trivially the case that the probability 
of an {\bf event} in $\cal{C}$ 
conditioned on the event $\Lambda$ is the probability of that event in the sCEG ${\cal{C}}_{\Lambda}$.

\subsection{Random variables on sCEGs}  

Random variables measurable with respect to the sigma field of $\mathcal{C}$ partition the set of atoms into events. So consider a random
variable $X$ with state space~$\mathbb{X}$, and let us denote the event that $X$ takes the value $x$ ($\in \mathbb{X}$) by $\Lambda_x$. Then the set
$\{ \Lambda_x \}_{x \in \mathbb{X}}$ partitions $\Lambda({\mathcal{C}})$. 

For any CEG there is a set of fairly transparent random variables
which includes as a subset the set of measurement variables of any BN-representation of the model, if such a representation exists. These are
called {\it cut-variables} and are discussed in detail in Section~3.2. In Figure~4 for example, we have a variable which could be called {\it symptom}, which
could take the values 1, 2 \& 3 (in some order) for routes traversing edges labelled {\it before}, {\it after} and {\it never}. These are not however 
the only variables we can define on a CEG, and we first consider some results for general variables.

\smallskip
Note that when we write $X \amalg Y$ we mean that $p(X = x, Y = y) =\break p(X = x)\ p(Y = y)\ \forall x \in {\mathbb{X}}, y \in {\mathbb{Y}}$, 
and that this is true for all distributions $P$ compatible with $\cal{C}$.
Now for an intrinsic event $\Lambda$, we can write $X \amalg Y\ |\ \Lambda$ if and only if $p(X = x, Y = y\ |\ \Lambda) = p(X = x\ |\ \Lambda)\ 
p(Y = y\ |\ \Lambda)$ for all values $x$ of $X$ and $y$ of $Y$ (see for example~\cite{DawidMilan}). That is $X \amalg Y\ |\ \Lambda\ \Leftrightarrow 
p(\Lambda_x, \Lambda_y\ |\ \Lambda) = p(\Lambda_x\ |\ \Lambda)\ p(\Lambda_y\ |\ \Lambda)$
for all $\Lambda_x \in \{ \Lambda_x \}_{x \in \mathbb{X}}$, $\Lambda_y \in \{ \Lambda_y \}_{y \in \mathbb{Y}}$.

\begin{lem} 	
\noindent{For a CEG $\cal{C}$, variables $X, Y$ measurable with respect to the sigma field of $\mathcal{C}$, and intrinsic conditioning event~$\Lambda$, 
the statement $X \amalg Y\ |\ \Lambda$ is true if and only if $X \amalg Y$ is true in the sCEG~${\cal{C}}_{\Lambda}$.}
\end{lem}

The proof of this lemma is in the appendix. 
This is a particularly useful property because it allows us to check any context-specific conditional independence property by checking a 
non-conditional independence property on a sub-sCEG.

To motivate the theory in the remainder of section~2 and in section~3, we need a bigger example.

\begin{ex}	
Our researcher from Example~1 now turns her attention to an ongoing study. Subjects who display the symptom {\rm S} (at any point) may
be given a drug, and the probability of receiving this drug is not dependent on their gender or when they displayed {\rm S}.
Those that develop the condition {\rm C} may be given treatment, and the probability of receiving this treatment is not dependent on their gender, 
when they displayed {\rm S}, or whether or not they received the earlier drug.
The CEG for this is given in Figure~5.

These two properties are depicted in the CEG by the positions $w_3$ \& $w_4$ being in the same stage, and by $w_{10}, w_{12}$ \& $w_{14}$ also being in the 
same stage. 

\begin{figure}[t]	
\includegraphics[width=12cm,height=9cm]{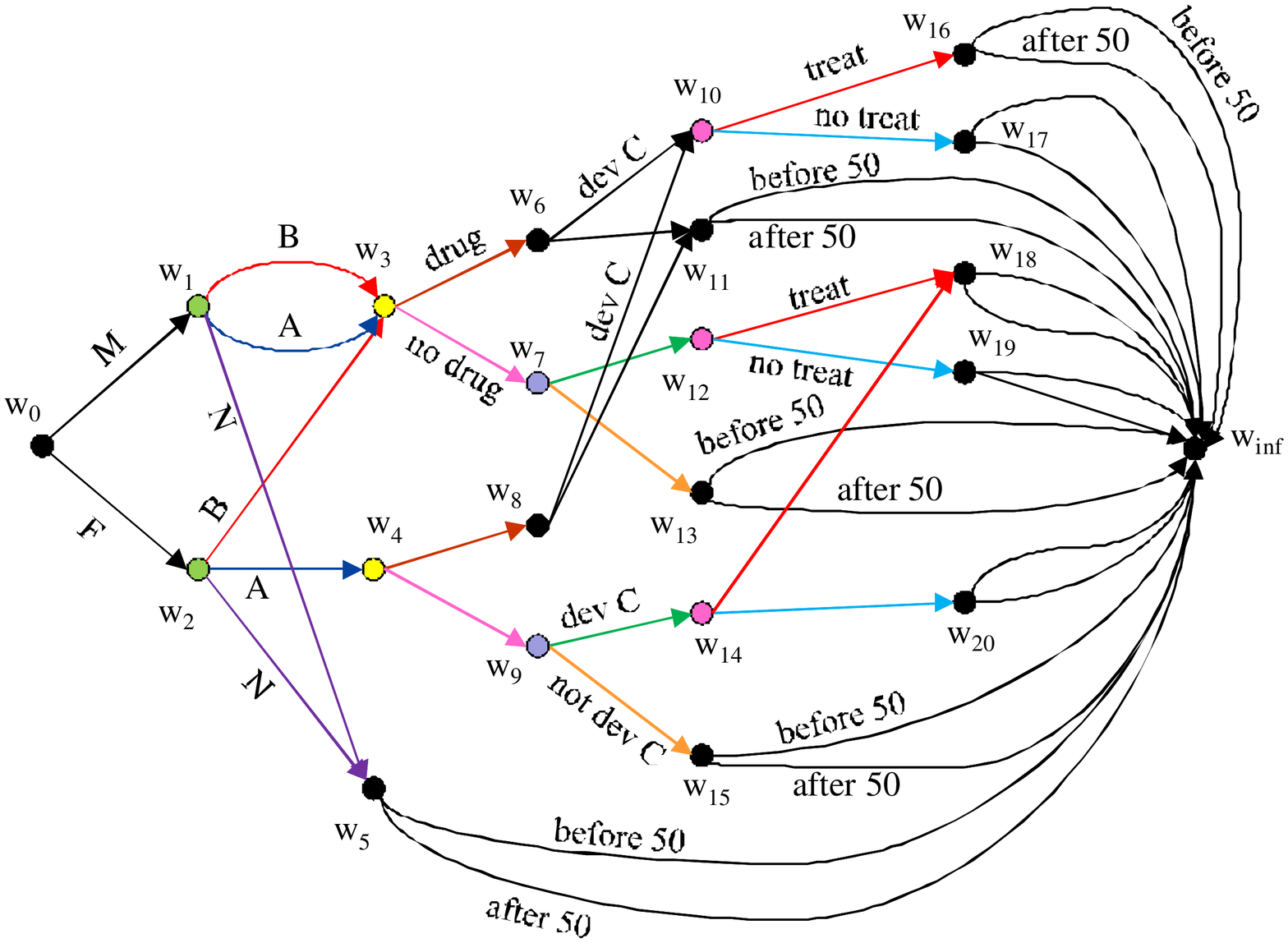}
\caption[]{CEG for Example 2}
\end{figure}

Figure~5 also tells us that treatment and life expectancy are independent of gender and when {\rm S} displayed, given both the event {\rm drug given \&
develop C} and the event {\rm drug given \& not develop C} (the positions $w_{10}$ \& $w_{11}$); and that life expectancy is independent of gender
and when {\rm S} displayed given the event {\rm drug not given, develop C \& treatment given} (the position $w_{18}$).


\smallskip
Our researcher is interested in the relationships between condition \& gender, and between condition \& when {\rm S} displayed, for the subgroups
(i) who were given the drug, and (ii) who displayed {\rm S} but were not given the drug.
\end{ex}

In BN-theory, if we wish to answer the query $X \amalg Y\ |\ Z$ ?, one way we might start doing this is by drawing the ancestral graph of 
$\{ X, Y, Z \}$ (see for example~\cite{Lauritzen2001}). We do this because variables in the BN which are not part of this graph have no influence 
on the outcome of our query.

There is no direct analogy for this graph in CEG-theory, but we can consider a pseudo-ancestral graph
associated with a set of events or variables. So in Example~2, all edges associated with treatment or life expectancy lie downstream
(ie. towards the sink-node) of the edges associated with gender, symptom, drug and condition, so we can simply curtail our CEG so that it does not
include these edges.

So in Figure~5, the positions $w_5, w_{10}, w_{11}, w_{12}, w_{13}, w_{14}$ \& $w_{15}$ are coalesced into a new sink-node $w_{\infty}$ as in
Figure~6. But $w_7$ \& $w_9$ in Figure~5 were in the same stage. As these nodes are now only one edge upstream of $w_{\infty}$, they get coalesced
into a single new position ($w_7$ in Figure~6). Notice how much simpler the pseudo-ancestral graph is than the original CEG.

\begin{figure}[t]	
\includegraphics[width=12cm,height=6.7cm,bbllx=54,bblly=150,bburx=777,bbury=555]{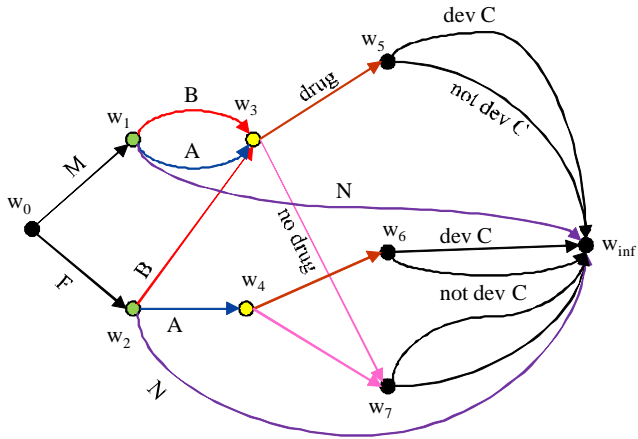}
\caption[]{Pseudo-ancestral graph for Example 2}
\end{figure}

We have noted above that stage-structure is often destroyed by conditioning on an intrinsic event, but that the set of sCEGs is closed under
this conditioning. So the remainder of our analysis is conducted on an uncoloured CEG.

\begin{figure}[t]	
\includegraphics[width=12cm,height=6.7cm,bbllx=54,bblly=150,bburx=777,bbury=555]{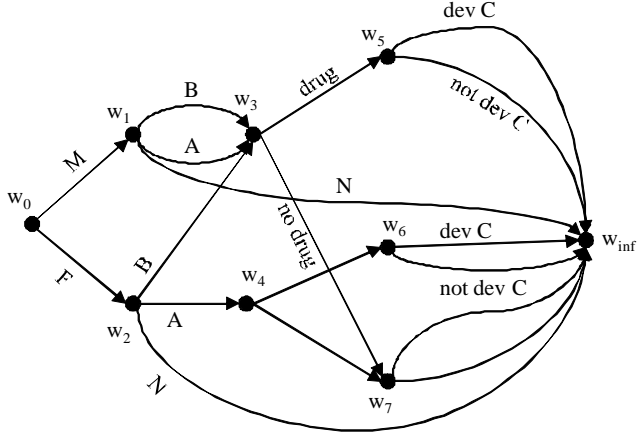}
\caption[]{Pseudo-ancestral sCEG $\cal{C}$ for Example 2}
\end{figure}

\smallskip
The graph in Figure~7 is the uncoloured pseudo-ancestral sCEG $\cal{C}$ associated with the queries that our researcher
is interested in. This graph is analogous to the moralised ancestral graph used in BN-based analysis.

There are two {\rm natural} variables which partition $\Lambda({\cal{C}})$ -- these are 
gender (which partitions $\Lambda({\cal{C}})$ into events which we will call {\rm M} (male) \& {\rm F} (female)), and symptom (which partitions 
$\Lambda({\cal{C}})$ into events which we will call {\rm B} ({\rm S} displayed before puberty), {\rm A} ({\rm S} displayed after puberty) \&
{\rm N} ({\rm S} never displayed)).

The variable associated with giving the drug partitions $\Lambda({\cal{C}})$ into three events -- drug given, {\rm S} displayed but drug not given, and 
{\rm S} not displayed and hence drug not given. As the third of these events is exactly the event {\rm N} above, we will for brevity describe the second
event (particularly when labelling edges) simply as {\rm drug not given} or {\rm no drug}.

The variable associated with condition {\rm C} partitions $\Lambda({\cal{C}})$ into three events -- {\rm C}~developed, {\rm S}~displayed but {\rm C} not 
developed, and {\rm S} not displayed and hence {\rm C}~not developed. Again, as the third of these events is exactly the event {\rm N}, we will for 
brevity describe the second event simply as {\rm C not developed}.

There is no ambiguity here as the queries our researcher is interested in correspond to conditioning on the events {\rm drug given}, and {\rm S
displayed but drug not given}. 

Her first question concerns the relationship between condition and when {\rm S} displayed for the subgroup who were 
given the drug. This requires conditioning on the event {\rm drug given}, so we draw the sub-sCEG ${\cal{C}}_{\Lambda}$ for this event. This is 
given in Figure~8.

\begin{figure}[t]	
\includegraphics[width=12cm,height=5.55cm,bbllx=54,bblly=220,bburx=777,bbury=555]{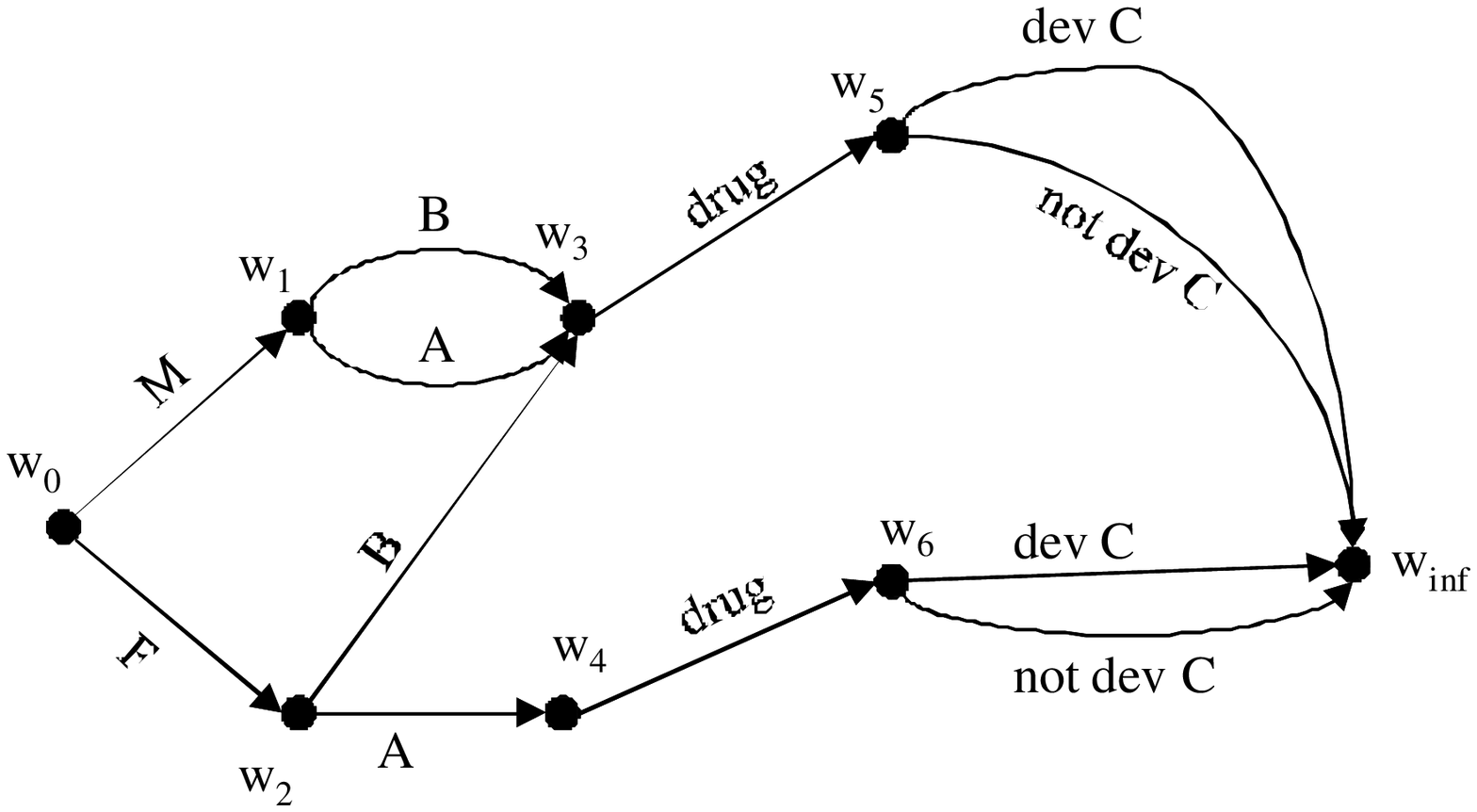}
\caption[]{${\cal{C}}_{\Lambda}$ for $\Lambda = \{ \hbox{drug given} \}$}
\end{figure}

The relationships she is interested in concern the probabilities $p_{\Lambda}(\hbox{\rm C developed}\ |\ \hbox{\rm B})$ and 
$p_{\Lambda}(\hbox{\rm C developed}\ |\ \hbox{\rm A})$, and these are given below:
\begin{align*}
p_{\Lambda}(&\hbox{\rm C developed}\ |\ \hbox{\rm B}) \\
	=\ &\{ p_{\Lambda}({\rm M})\ p_{\Lambda}({\rm B}\ |\ {\rm M})\times 1 \times p_{\Lambda}(\hbox{C developed}\ |\ ({\rm M}, {\rm B})\ {\rm or}\ 
		({\rm M}, {\rm A})\ {\rm or} \ ({\rm F}, {\rm B})) \\
	&+ p_{\Lambda}({\rm F})\ p_{\Lambda}({\rm B}\ |\ {\rm F})\times 1 \times p_{\Lambda}(\hbox{C developed}\ |\ ({\rm M}, {\rm B})\ {\rm or}\ 
		({\rm M}, {\rm A})\ {\rm or} \ ({\rm F}, {\rm B})) \}  \\
	& \div \{ p_{\Lambda}({\rm M})\ p_{\Lambda}({\rm B}\ |\ {\rm M}) + p_{\Lambda}({\rm F})\ p_{\Lambda}({\rm B}\ |\ {\rm F}) \}  \\ 
	=\ &p_{\Lambda}(\hbox{C developed}\ |\ ({\rm M}, {\rm B})\ {\rm or}\ ({\rm M}, {\rm A})\ {\rm or}\ ({\rm F}, {\rm B}))  \\
	=\ &p (\hbox{C developed}\ |\ (({\rm M}, {\rm B})\ {\rm or}\ ({\rm M}, {\rm A})\ {\rm or}\ ({\rm F}, {\rm B})),\ \hbox{drug given})  &(2.1)
\end{align*}
\noindent{Note:}
\begin{enumerate}
\item We do not need for our purposes here to evaluate the $p_{\Lambda}(\dots)$ probabilities, but if we wished to we could use the expression from
	Theorem~1.
\item The expression (2.1) is still the simplest expression even if we were to reintroduce stage-structure and let $w_1$ \& $w_2$ be in the same
	stage.
\end{enumerate}
\begin{align*}
p_{\Lambda}(&\hbox{\rm C developed}\ |\ \hbox{\rm A}) \\
	=\ &\{ p_{\Lambda}({\rm M})\ p_{\Lambda}({\rm A}\ |\ {\rm M})\times 1 \times p_{\Lambda}(\hbox{C developed}\ |\ ({\rm M}, {\rm B})\ {\rm or}\ 
		({\rm M}, {\rm A})\ {\rm or} \ ({\rm F}, {\rm B})) \\
	&+ p_{\Lambda}({\rm F})\ p_{\Lambda}({\rm A}\ |\ {\rm F})\times 1 \times p_{\Lambda}(\hbox{C developed}\ |\ {\rm F}, {\rm A}) \}  \\
	& \div \{ p_{\Lambda}({\rm M})\ p_{\Lambda}({\rm A}\ |\ {\rm M}) + p_{\Lambda}({\rm F})\ p_{\Lambda}({\rm A}\ |\ {\rm F}) \}  
\end{align*}
\noindent{which clearly does not equal expression~(2.1).}

\begin{itemize}
\item[3.] The denominator is of course $p_{\Lambda}({\rm A})$, but even if we let $w_1$ \& $w_2$ be in the same stage, the above expression 
	only simplifies to\hfill\break $p_{\Lambda}({\rm M})\ p_{\Lambda}(\hbox{C developed}\ |\ ({\rm M}, {\rm B})\ {\rm or}\ ({\rm M}, {\rm A})\ 
	{\rm or}\ ({\rm F}, {\rm B})) + 
	p_{\Lambda}({\rm F})\ p_{\Lambda}(\hbox{C developed}\ |\ {\rm F}, {\rm A})$, which still does
not equal expression~(2.1). 
\end{itemize}

Suppose we now consider the subgroup who displayed S but were not given the drug and the sub-sCEG~${\cal{C}}_{\Lambda}$ for the event {\rm drug not given}. 
This CEG is given in Figure~9.


The corresponding probabilities are:
\begin{align*}
p_{\Lambda}(\hbox{\rm C developed}\ |\ &\hbox{\rm B}) \\
	=\ &\{ p_{\Lambda}({\rm M})\ p_{\Lambda}({\rm B}\ |\ {\rm M})\times 1 \times p_{\Lambda}(\hbox{C developed}) \\
	&+ p_{\Lambda}({\rm F})\ p_{\Lambda}({\rm B}\ |\ {\rm F})\times 1 \times p_{\Lambda}(\hbox{C developed}) \}  \\
	& \div \{ p_{\Lambda}({\rm M})\ p_{\Lambda}({\rm B}\ |\ {\rm M}) + p_{\Lambda}({\rm F})\ p_{\Lambda}({\rm B}\ |\ {\rm F}) \}  \\ 
	=\ &p_{\Lambda}(\hbox{C developed}) \quad = p (\hbox{C developed}\ |\ \hbox{drug not given}) 
\end{align*}
\begin{align*}
p_{\Lambda}(\hbox{\rm C developed}\ |\ &\hbox{\rm A}) \\
	=\ &\{ p_{\Lambda}({\rm M})\ p_{\Lambda}({\rm A}\ |\ {\rm M})\times 1 \times p_{\Lambda}(\hbox{C developed}) \\
	&+ p_{\Lambda}({\rm F})\ p_{\Lambda}({\rm A}\ |\ {\rm F})\times 1 \times p_{\Lambda}(\hbox{C developed}) \}  \\
	& \div \{ p_{\Lambda}({\rm M})\ p_{\Lambda}({\rm A}\ |\ {\rm M}) + p_{\Lambda}({\rm F})\ p_{\Lambda}({\rm A}\ |\ {\rm F}) \}  \\ 
	=\ &p_{\Lambda}(\hbox{C developed}) \quad = p (\hbox{C developed}\ |\ \hbox{drug not given})  
\end{align*}
So we have that {\it whether {\rm C} developed is independent of when {\rm S}~displayed, given that {\rm S}~was displayed but the drug was not given}.

\begin{figure}[t]	
\includegraphics[width=12cm,height=5.5cm,bbllx=54,bblly=170,bburx=777,bbury=500]{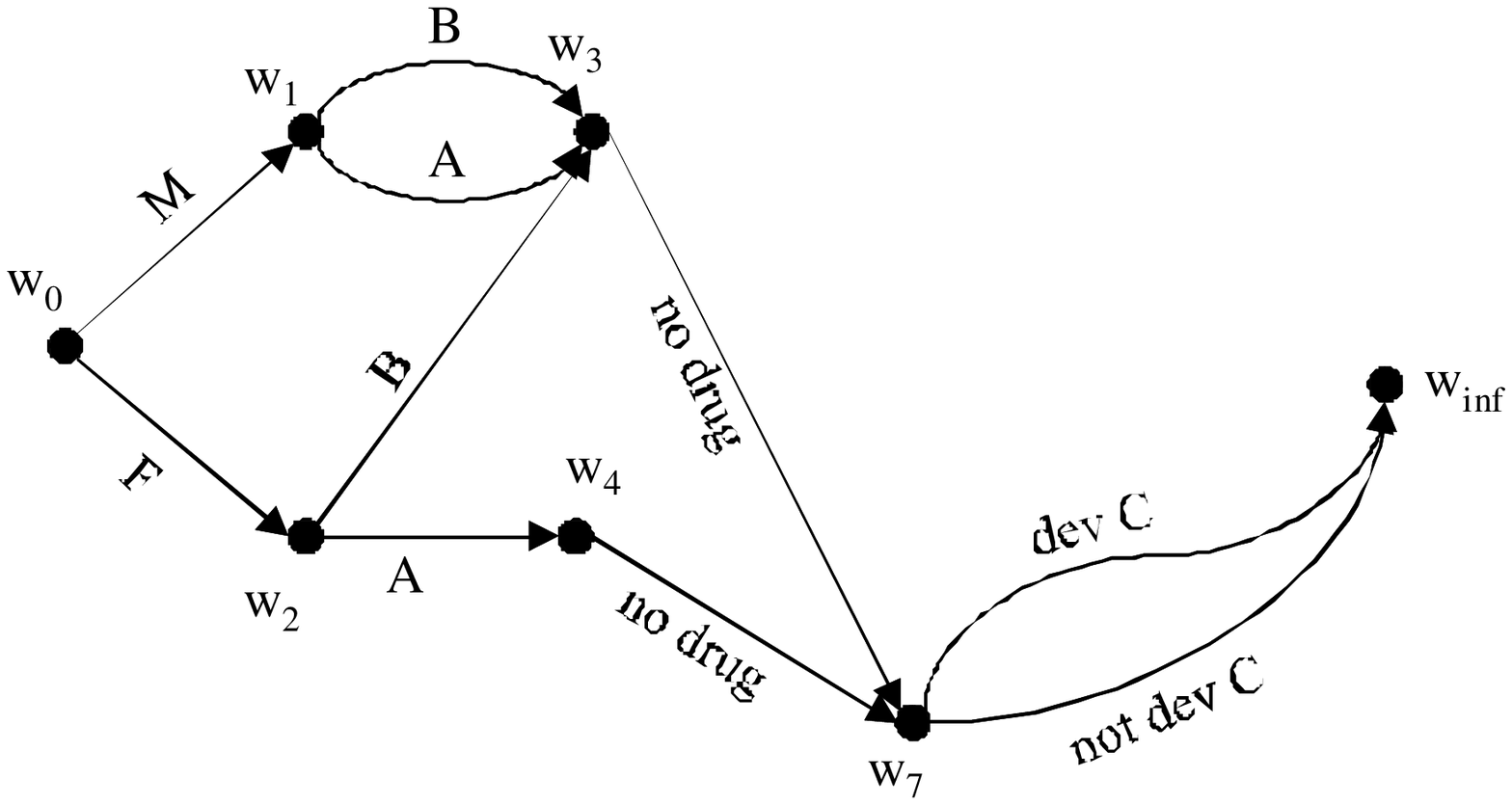}
\caption[]{${\cal{C}}_{\Lambda}$ for $\Lambda = \{ \hbox{drug not given} \}$}
\end{figure}

\begin{itemize}
\item[4.] We do not need to consider the case where S never displayed, as this has no intersection with $\Lambda$: Given that S {\it displayed
but drug not given}, I {\bf know} that S was displayed, but further knowledge of when it was displayed is irrelevant for prediction of whether
or not the subject developed C,
\item[5.] This context-specific conditional independence property holds whether or not we reintroduce stage-structure and let $w_1$ \& $w_2$
be in the same stage.
\end{itemize}

Let us now consider our researcher's other queries to do with the relationship between condition and gender for our subgroups. From Figure~8
we can see that if $\Lambda = \{\hbox{drug given}\}$ then:
\begin{align*}
p_{\Lambda}(&\hbox{\rm C developed}\ |\ \hbox{\rm M}) \\
	=\ &p_{\Lambda}(\hbox{A or B}\ |\ {\rm M})\times 1 \times p_{\Lambda}(\hbox{C developed}\ |\ ({\rm M}, {\rm B})\ {\rm or}\ ({\rm M}, {\rm A})\ 
			{\rm or} \ ({\rm F}, {\rm B}))    &(2.2)
\end{align*}
\noindent{and $p_{\Lambda}(\hbox{A or B}\ |\ {\rm M}) = 1$, since A \& B are the only edges leaving $w_1$ in~${\cal{C}}_{\Lambda}$.} 
\begin{align*}
p_{\Lambda}(\hbox{\rm C developed}\ &|\ \hbox{\rm F}) \\
	=\ &p_{\Lambda}({\rm B}\ |\ {\rm F})\times 1 \times p_{\Lambda}(\hbox{C developed}\ |\ ({\rm M}, {\rm B})\ {\rm or}\ ({\rm M}, {\rm A})\ {\rm or} 
				\ ({\rm F}, {\rm B}))    \\
	&+ p_{\Lambda}({\rm A}\ |\ {\rm F})\times 1 \times p_{\Lambda}(\hbox{C developed}\ |\ {\rm F}, {\rm A})
\end{align*}
\noindent{which clearly does not equal expression~(2.2), and this is true even if we reintroduce  stage-structure and let $w_1$ \& $w_2$
be in the same stage.}

\smallskip
From Figure~9 we can see that if $\Lambda = \{\hbox{drug not given}\}$ then:
\begin{align*}
p_{\Lambda}(\hbox{\rm C developed}\ |\ &\hbox{\rm M}) \\
	=\ &p_{\Lambda}(\hbox{A or B}\ |\ {\rm M})\times 1 \times p_{\Lambda}(\hbox{C developed})   \\
	=\ &p_{\Lambda}(\hbox{C developed}) \quad =  p(\hbox{C developed}\ |\ \hbox{drug not given})  \\
p_{\Lambda}(\hbox{\rm C developed}\ |\ &\hbox{\rm F}) \\
	=\ &p_{\Lambda}({\rm B}\ |\ {\rm F})\times 1 \times p_{\Lambda}(\hbox{C developed})   \\
	&+ p_{\Lambda}({\rm A}\ |\ {\rm F})\times 1 \times p_{\Lambda}(\hbox{C developed})   \\
	=\ &p_{\Lambda}(\hbox{C developed}) \quad =   p(\hbox{C developed}\ |\ \hbox{drug not given}) 
\end{align*}
So we have that {\it whether {\rm C} developed is independent of gender, given that {\rm S}~was displayed, but the drug was not given}, and this is true
irrespective of whether we reintroduce stage-structure.

\bigskip
We notice that the topological feature which distinguishes Figure~9 from Figure~8 is that in Figure~9 there is a cut-vertex 
(a single vertex not $w_0$ or $w_{\infty}$, through which all routes in the graph pass) lying between the edges
associated with gender \& symptom (upstream) and those associated with condition (downstream). We return to cut-vertices and to their role
in independence queries in section~3.


Note also that the example above gives ample justification for working with sCEGs when considering conditional independence queries, rather
than their coloured counterparts.

\section{A separation theorem for simple CEGs}		

In section 2.7 we introduced random variables on sCEGs. In section~3.1 we develop this idea, before providing a separation theorem for sCEGs
in section~3.2.

\subsection{Position variables}		

As noted in Section~1, modified BNs of one type or another are widely used because real problems tend to contain more symmetries than can be
represented by a standard BN. What is generally not addressed in papers on these types of graphs is the consequence that this extra structure
has for the Markov relationships between the problem variables. With CEGs we can address this explicitly \& automatically, and the first step
towards doing this is to consider model variables which are more fundamental than the measurement variables customarily considered when working
with BNs. So in
this section we describe two types of elementary random variables, measurable with respect to the sigma field of~$\mathcal{C}$, that can be 
identified with each position $w \in V(\mathcal{C}) \backslash \{ w_{\infty} \}$. These are the variables 
$\{ I(w) \} \}$ and $\{ X(w) \} $  defined 
below.

Note that when we say that a variable $X$ takes the value $x$, this is equivalent to saying that an individual from our population has a development
which we equate with a route $\lambda$, and that this route $\lambda$ is an element of $\Lambda_x$, the event corresponding to $X = x$.

For a position $w$, $I(w)$ can take the values 1 or 0 depending on whether this individual is on a route $\lambda$ which does or does not pass through $w$.
So: 
$$I(w)=\left\{ 
\begin{array}{cl} 
1 & \hbox{if}\ w \in \lambda \\
0 & \hbox{if}\ w \not\in \lambda
\end{array}
\right.
$$
\noindent{(where as above, $w \in \lambda$ means that the position $w$ lies on the route $\lambda$).}

\smallskip
Up until now we have labelled edges by their start and endpoints (eg. $e(w, w')$), but we can also label the edges leaving a position $w$ by a set
of arbitrary labels of the form $e_x(w)$ ($x = 1, 2, \dots$). We define $X(w)$ by:
$$X(w)=\left\{ 
\begin{array}{cl} 
x & \hbox{if}\ e_x(w) \in \lambda \\
0 & \hbox{if}\ w \not\in \lambda 
\end{array}
\right.
$$
So $X(w) = x\ (\ne 0)$ if our individual is on a route $\lambda$ which passes through $w$ and the edge $e_x(w)$.

Recall that a CEG depicts all possible histories of a unit in a population, and gives a probability
distribution over these histories. However, when a single unit traverses one of the routes in the CEG, values are
assigned to $I(w)$ \& $X(w)$ for all positions $w \in V({\mathcal{C}})$.

Notice that since $I(w)$ is clearly a function of $X(w)$, to specify a full joint distribution over the position variables, it is sufficient to 
specify the joint distribution of $\{ X(w) : w \in V(\mathcal{C}) \backslash \{ w_{\infty} \} \}$. Note also that all atoms $\lambda$ can be 
expressed as an intersection:
\begin{align*}
&\lambda =\bigcap\limits_{w \in \lambda }\left\{ X(w)=x_{\lambda}\right\}, \\
\intertext{and events in the sigma algebra of $\cal{C}$ as the union of these atoms:} 
&\Lambda = \bigcup\limits_{\lambda \in \Lambda}\left\{ \bigcap\limits_{w\in \lambda }\left\{ X(w)=x_{\lambda}\right\}
\right\},
\end{align*}
\noindent{where $x_{\lambda}$ ($\ne 0$) is the unique value of $X(w)$ labelling the edge in the route $\lambda$.}

Up until this point we have used the words {\it upstream} and {\it downstream} rather loosely -- in the context of {\bf sets} of {\bf edges} we
have simply used these words to mean further towards $w_0$ and further towards $w_{\infty}$; but we need to formalise the meanings here in the
context of {\bf positions}. So when we say that $w_1$ is upstream of $w_2$, or $w_2$ is downstream of $w_1$, we mean that $w_1 \prec w_2$.

For any set $A \subset V({\mathcal{C}})$, let $\boldsymbol{X}_{A}$ denote the set of
random variables $\{ X(w):\break w\in A \}$ and $\boldsymbol{I}_{A}$ the set $\left\{ I(w):w\in A\right\}$.
Also, for any $w \in V({\mathcal{C}})$, let $U(w)$ be the set of positions in $V({\mathcal{C}})$ which lie 
upstream of the position $w$, $D(w)$ the set of positions which lie downstream of $w$, $U^c(w)$
the set of positions which do not lie upstream of $w$, and $D^c(w)$ the set of positions which do not lie 
downstream of $w$.

\begin{lem}   
For any sCEG $\cal{C}$ and position $w \in V({\mathcal{C}}) \backslash \{ w_{\infty} \}$, the variables  
$I(w), X(w)$ exhibit the {\rm position independence property} that
$$X(w)\amalg \boldsymbol{X}_{D^c(w)}\ |\ I(w)$$
\end{lem}

This result (an extension of the Limited Memory Lemma of~\cite{newcap}) is analogous to the Directed Markov 
property which can be used to define BNs (see for example~\cite{Pearl2000}), and 
which states that a BN vertex-variable is independent of its non-descendants given its parents. It provides a 
set of conditional independence statements that can simply be read from the graph, one for each position 
in~$V({\mathcal{C}})$. The proof of the lemma is in the appendix.

The statement that $X(w)\amalg \boldsymbol{X}_{D^c(w)}\ |\ (I(w) = 1)$ can be read as: \emph{Given a unit reaches 
a position $w\in V(\mathcal{C})$, whatever happens immediately after $w$ is independent of not only all developments
through which that position was reached, but also of all positions that
logically have not happened or could not now happen because the unit has passed
through $w$.} 

\subsection{Theorem and corollaries} 		

It is doubtful whether BNs would have enjoyed their enormous popularity if it were not so apparently easy to read conditional independence
properties from them. In particular, the existence of the d-separation theorem~\cite{Laurdsep,VandP} has allowed all practitioners to make some
attempt at model interpretation with some degree of confidence.

The presence of any context-specific conditional independence structure however severely hampers analysts using BNs in their attempts to get
accurate pictures of the structure of their problems~\cite{kollerUAI1996,PooleandZ}. In earlier sections of this paper (and in particular in 
section~2.7) we have been
developing the theory needed for reading and representing (context-specific) conditional independence structure using CEGs. In particular, 
Lemma~1 allows us to consider context-specific queries by looking at the relevant sub-CEG; and Example~2 provides the rationale for looking
at sCEGs. We now provide a separation theorem for sCEGs.

\smallskip
Using the standard terminology of non-probabilistic graph theory, we call a position $w \in V({\cal{C}}) \backslash \{w_{\infty} \}$ a {\em cut-vertex} 
if the removal of $w$ and its associated edges from $\cal{C}$ would result in a graph with two disconnected components. An alternative description
would be {\em a position other than $w_0$ through which all routes pass}. 
We also remind readers at this point that when we write (for example) $X \amalg Y$ we mean 
that $p(X = x, Y = y) = p(X = x)\ p(Y = y)\break \forall x \in {\mathbb{X}}, y \in {\mathbb{Y}}$, and that this is true for all distributions $P$
compatible with $\cal{C}$.

\begin{thm}  
In an sCEG $\cal{C}$ with $w_1, w_2 \in V(C) \backslash \{w_{\infty}\}$ and $w_2 \not\prec w_1$,\break $X(w_1) \amalg X(w_2)$ if and only if
either (i) there exists a cut-vertex $w$ such that $w_1 \prec w \prec w_2$, or (ii) $w_2$ is itself a cut-vertex. 
\end{thm}

The proof of this theorem is in the appendix.
The variables $\{ I(w) \} \}$ and $\{ X(w) \}$ have an obvious intrinsic mathematical interest, but for more practical purposes we need to be able to make 
statements about the relationships between variables which are more closely analogous to the measurement variables used in BN-based analysis. So, in
the same way that our primitive probabilities were used to build probabilities of subpaths and routes, we can use the $X(w)$ variables to build
new {\it bigger} variables which have a more transparent interpretation for the analyst.

\smallskip
In Figure~4, let $X(w_i)$ (for $i = 5,6,7,8,9$) equal 1 if an individual has a development which takes them through the position $w_i$ and then they
die before the age of 50, and equal 2 if they have a development which takes them through the position $w_i$ and then they die after the age of 50.
Since an individual's development will take them through one \& only one of $\{ w_5, w_6, w_7, w_8, w_9 \}$, we can define a life expectancy
indicator across the whole CEG by
$$\sup_{w_i : i \in \{5,6,7,8,9\} } X(w_i)$$
which takes the value 1 if an individual dies before the age of 50, or 2 if they die after the age of 50.

\smallskip
Analogously with the idea of a cut-vertex, a {\it position cut} is a set of positions the removal of which from $V({\cal{C}})$ would result in a graph
with two disconnected components. This is formalised in Definition~6.

\begin{defn}	
{\bf Position cut.} A set of positions $W \subset V({\cal{C}}) \backslash \{w_0, w_{\infty}\}$ is a {\em position cut} if $\{ \Lambda(w) : w \in W \}$
forms a partition of $\Lambda({\cal{C}})$.
\end{defn}

\noindent{As noted above, for any position cut $W$, we can define a {\it cut-variable}; this is formalised in Definition~7.}

\begin{defn}	
{\bf Cut-variable.} For a position cut $W$, the random variable $X(W) \equiv \sup_{w \in W} X(w)$ is called a {\em cut-variable}.
\end{defn}

\noindent{Note that $X(W)$ can also be defined as $X(W) \equiv \sum_{w \in W} X(w)$.
The equivalence of the two forms comes from the fact that $X(w) > 0$ for one \&~only one position $w \in W$.}

\smallskip
In Figure~4 we have the obvious cut-variables {\it gender} and {\it symptom}. If we assign values of 1 to edges labelled {\it develop} C, 2 to 
{\it not develop} C, 3 to {\it die before 50}, and 4 to {\it die after 50}, then
$X(W)$ for $W = \{ w_3, w_4, w_5 \}$ becomes a more sophisticated cut-variable for developing the condition: $X(W)$ takes the value 1 if \& only if an 
individual develops C, but $X(W) = 2$ tells us that an individual displayed symptom S yet did not develop C, and $X(W) = 3$ or 4 tells us that an
individual did not display S and therefore did not develop C.

\smallskip
Theorem~2 allows us to look at the detail of the Markov structure depicted by our CEGs. The following corollaries allow us to get a broader picture.

\begin{cor}	
For an sCEG $\cal{C}$ with position cuts $W_a$ and $W_b$, the property\break $X(w_1) \amalg X(w_2)$ holding for any $w_1 \in W_a,\ w_2 \in W_b$
implies that  $X(W_a) \amalg X(W_b)$.
\end{cor}

So, as one might expect, the presence of a cut-vertex in an sCEG renders cut-variables upstream of this vertex independent of cut-variables downstream
of the vertex.
The proof of the corollary is in the appendix.

As already noted, CEGs have been designed for the representation and analysis of asymmetric problems; and for symmetric problems a graph such as
a BN is more appropriate. But it is clear that where a problem can also be adequately represented as a BN (without too much context-specific structure),
the set of cut-variables of a CEG-representation must contain the set of variables associated with the vertices of the BN, as these are simply the
measurement variables of the problem. Hence, if an sCEG $\cal{C}$ represents a model which admits a product space structure, $M, N$ are
measurement variables of the model associated with position cuts $W_M, W_N$, then the property $M \amalg N$ holds providing that
$X(w_m) \amalg X(w_n)$ for any $w_m \in W_M, w_n \in W_N$. This result follows immediately from Corollary~1.

Of more interest to analysts of asymmetric problems is the result given in Corollary~2, which ties together the ideas presented in Corollary~1
and Lemma~1.

\begin{cor}	
Let $\cal{C}$ be a CEG with position cuts $W_a, W_b$, and $\Lambda$ an event intrinsic to $\cal{C}$. If, in the sCEG ${\cal{C}}_{\Lambda}$, 
there exists a cut-vertex $w$ such that\break $W_a \prec w \prec W_b$, then $X(W_a) \amalg X(W_b)\ |\ \Lambda$.
\end{cor}

The proof of this corollary is in the appendix. We can immediately deduce that if a CEG $\cal{C}$ represents a model which admits a product space
structure, $M, N$ are measurement variables of the model associated with position cuts $W_M, W_N$, and $\Lambda$ is an event intrinsic to~$\cal{C}$,
then if in the sCEG ${\cal{C}}_{\Lambda}$ there exists a cut-vertex $w$ such that $W_M \prec w \prec W_N$, the property $M \amalg N\ |\ \Lambda$
must hold.


Recall from Section~2.7 that for a measurement variable $X$ with state space~${\mathbb{X}}$, the event that $X$ takes the value $x$ ($\in {\mathbb{X}}$)
is denoted by $\Lambda_x$, and the set $\{ \Lambda_x \}_{x \in {\mathbb{X}}}$ partitions $\Lambda({\cal{C}})$. So the query $M \amalg N\ |\ X$ ? 
can be answered by checking the queries $M \amalg N\ |\ \Lambda_x$ ? for each $x \in {\mathbb{X}}$.
If our problem elicitation indicates that there are no context-specific variations in independence properties connected with conditioning on the
variable $X$, we can answer the query $M \amalg N\ |\ X$ ? by looking at a single graph ${\cal{C}}_{\Lambda_x}$ for some convenient value $X = x$.

Moreover, although this argument has been constructed under the assumptions that $\cal{C}$ admits a product space structure, and that $M, N$ \& $X$
are measurement variables of the problem, these assumptions are not strictly necessary; it is sufficient that $M$ \& $N$ are cut-variables, and
that $\{ \Lambda_x \}_{x \in {\mathbb{X}}}$ partitions $\Lambda({\cal{C}})$. And even these conditions can be relaxed, as we see in Example~3.

\begin{ex}	
An alternative drug becomes available, resulting in a revised sCEG as in Figure~10.
\end{ex}

\begin{figure}[t]	
\includegraphics[width=12cm,height=7.4cm,bbllx=54,bblly=150,bburx=777,bbury=597]{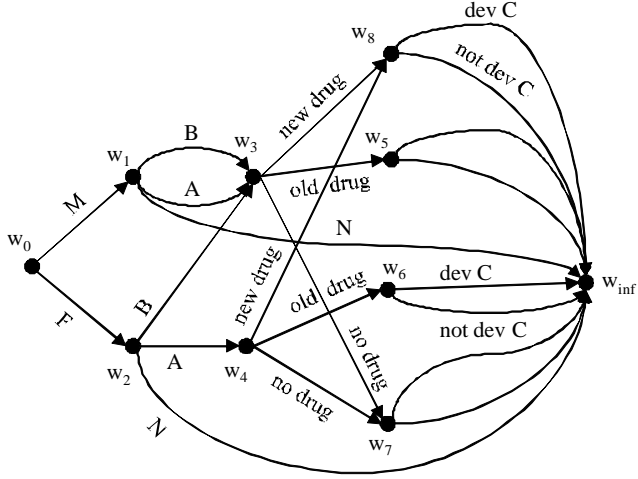}
\caption[]{sCEG for Example 3}
\end{figure}

Let $W_a = \{w_0\}, W_b = \{w_1, w_2\}, W_c = \{w_3, w_4\}$ and $W_d = \{w_5, w_6, w_7, w_8\}$. Now, unlike $W_b$, the sets $W_c$ \& $W_d$ are not 
position-cuts as they do not partition~$\Lambda({\cal{C}})$. However, we can still define
\begin{align*}
X(W_c) = \sup_{w \in W_c} X(w), \\
X(W_d) = \sup_{w \in W_d} X(w)
\end{align*}
$X(W_c), X(W_d)$ (although not cut-variables) are both measurable with respect to the sigma field of $\cal{C}$, but can, unlike $X(W_a)$ or $X(W_b)$, 
take zero values, if a patient does not display the symptom.

\noindent{If we let $\Lambda_1$ be the event S {\it displayed but drug not given}, then we get the sub-SCEG ${\cal{C}}_{\Lambda_1}$ shown in Figure~9, from which 
we can read the statement}
$$(X(W_a), X(W_b)) \amalg X(W_d)\ |\ \Lambda_1.$$
If we let $\Lambda_2$ be the event {\it old drug given}, then we get the graph ${\cal{C}}_{\Lambda_2}$ shown in Figure~8, and as we have already shown,
$$(X(W_a), X(W_b)) \; {\hbox{{\bf /}\llap{$\amalg$}}}\; X(W_d)\ |\ \Lambda_2.$$
If we let $\Lambda_3$ be the event {\it new drug given}, then we get a graph ${\cal{C}}_{\Lambda_3}$ which differs from that in Figure~9 only in that
the cut-vertex is now $w_8$, not $w_7$. We can then read the statement
$$(X(W_a), X(W_b)) \amalg X(W_d)\ |\ \Lambda_3.$$
Note that $\{ \Lambda_i \}_{i = 1,2,3}$ here does not partition $\Lambda({\cal{C}})$.

Clearly we can call $X(W_a)$ \& $X(W_b)$ {\it gender} ($X_G$) \& {\it symptom} ($X_S$).
If we let $X(w_3)$ \& $X(w_4)$ take the values 1, 2 \& 3 for the outcomes {\it no drug}, {\it old drug} \& {\it new drug}, then $X(W_c)$ takes the values
0, 1, 2 \& 3 for {\it did not display} S {\it so did not receive drug}, {\it displayed} S {\it but did not receive drug}, {\it received old drug} and
{\it received new drug}. So there is also no ambiguity in calling $X(W_c)$ {\it drug} ($X_D$). Taking a similar approach to
$\{X(w_i)\}_{i \in \{5,6,7,8\}}$ we find that there is also no ambiguity in calling $X(W_d)$ {\it condition} ($X_C$), and (since $X(W_c) = 0 
\Rightarrow X(W_d) = 0$) collecting these statements together gives the property
$$X_C \amalg (X_G, X_S)\ |\ (X_D \ne 2),$$
ie. {\it condition} is independent of {\it gender} \& {\it symptom} given that did not receive the old drug.

\section{Discussion} 
 
Chain Event Graphs were introduced for the representation \& analysis of problems for which the use of Bayesian Networks is not ideal.
The class of models expressible as a CEG includes as a proper subset the class
of models expressible as faithful regular or context-specific BNs on finite variables. Unlike the~BN, the CEG embodies
the structure of the model state space and any context-specific information in its topology. In this paper we have justified the use of sCEGs for 
investigating context-specific conditional independence queries of the form $X \amalg Y\ |\ \Lambda\ $ ?,
and provided a separation theorem for sCEGs and position variables. The introduction of cut-variables 
(analogous to BN measurement variables, but more flexible) provides a repertoire of techniques which will
enable researchers to tackle a comprehensive collection of conditional independence enquiries on models of asymmetric
problems for which the available quantitative dependence information cannot all be embodied in the DAG of a~BN.

The research that led to this paper also yieded a number of other questions, some of which are discussed here. The most obvious of these is {\it Does
the {\rm only if} part of Theorem~2 hold if we allow constraints on a CEG's edge-probabilities such as two edge-probabilities being equal?} The
short answer is {\it No}, but the problem is somewhat more subtle than this answer suggests. Some preliminary work on this is described 
in~\cite{Crismsep}, but a more comprehensive analysis awaits a future paper.

For illustrative convenience the CEGs in the examples in this paper have been constructed in temporal order, but this is not the only valid
ordering of a CEG. In~\cite{CausalAI} for instance, we had a CEG representing a police investigation where the order of events was that in which
the police took action or discovered evidence (Extensive Form order~\cite{JandPDTs}). At the simplest level, there are valid reorderings of a CEG in
which the cut-variables appear in a different sequence, and there is a set of rules governing when adjacent cut-variables can be {\it swapped} to 
produce a different valid ordering. For CEGs depicting models which have a natural product space structure with no context-specific anomolies, 
these rules are relatively straightforward, but for more general CEGs where we might need to consider swaps of sets of adjacent edges rather
than of cut-variables, the rules become very complex. However, it seems fairly certain that if two cut-variables in a coloured CEG are independent
then there is a valid reordered pseudo-ancestral CEG of these variables in which the variables are separated by a cut-vertex. We hope to yield 
more light on this in a future paper.

In \cite{Lornaetal} we have also looked at infinite CEGs where an individual might come back to (essentially) the same state at some future time
point. These problems can be expressed as a CEG analogous to the 2-time-sliced Dynamic BN~\cite{KorbandN}, or as a graph which is no longer acyclic.
Both representations involve modification to the rules governing conditional independence structure. This is discussed in~\cite{Lornaetal}, but
there is an opportunity here for developing CEG semantics further.

\section*{Appendix: Proofs}

\noindent{\bf Proof of Theorem~1:}

\smallskip
\noindent{Consider the underlying tree $\cal{T}$ of the CEG ${\cal{C}}({\cal{T}})$. The event $\Lambda$ corresponds to a union of routes of
$\cal{T}$. Let ${\cal{T}}_{\Lambda}$ be the reduced tree consisting only of the vertices, edges \& routes that comprise $\Lambda$.

If we denote the probabilities of events in ${\cal{T}}_{\Lambda}$ by $p_{\Lambda}(..)$, then clearly we require that 
$p_{\Lambda}(\lambda) = p(\lambda\ |\ \Lambda)$.

Once route probabilities in a tree are given, edge-probabilities are uniquely defined. So letting the edge-probabilities in ${\cal{T}}_{\Lambda}$
be denoted by $\hat{\pi}_e (v'\ |\ v)$, and letting the route $\lambda \in \Lambda$ be described by its edges as:
$$\lambda = \Lambda(e(v_0, v_1)) \cap \Lambda(e(v_1, v_2)) \cap \dots \cap \Lambda(e(v_p, v_q)),$$
we have:
\begin{align*}
p_{\Lambda}(\lambda) =\ &p(\lambda\ |\ \Lambda) \\
	=\ &p(\Lambda(e(v_0, v_1)), \Lambda(e(v_1, v_2)), \dots \Lambda(e(v_p, v_q))\ |\ \Lambda) \\
	=\ &p(\Lambda(e(v_p, v_q))\ |\ \Lambda(e(v_0, v_1)), \Lambda(e(v_1, v_2)), \dots , \Lambda) \\
	&\times \dots \times p(\Lambda(e(v_1, v_2))\ |\ \Lambda(e(v_0, v_1)), \Lambda) \\	
	&\times p(\Lambda(e(v_0, v_1))\ |\ \Lambda) \\
	=\ &p_{\Lambda} (\Lambda(e(v_p, v_q))\ |\ \Lambda(e(v_0, v_1)), \Lambda(e(v_1, v_2)), \dots) \\
	&\times \dots \times p_{\Lambda} (\Lambda(e(v_0, v_1))) \\
	=\ &p_{\Lambda} (\Lambda(e(v_p, v_q))\ |\ \Lambda(v_p)) \\
	&\times \dots \times p_{\Lambda} (\Lambda(e(v_1, v_2))\ |\ \Lambda(v_1)) \\	
	&\times p_{\Lambda} (\Lambda(e(v_0, v_1))) \\
\intertext{using the Markov property of trees from section~2.2}
	=\ &\prod_{e(v, v') \in \lambda} p_{\Lambda}(\Lambda(e(v, v'))\ |\ \Lambda(v)) \\
	=\ &\prod_{e(v, v') \in \lambda} \hat{\pi}_e (v'\ |\ v)
\end{align*}
\noindent{If we now let ${\cal{C}}_{\Lambda}$ inherit the edge-probabilities from ${\cal{T}}_{\Lambda}$, we have:}
\begin{align*}
p_{\Lambda}(\lambda) =\ &\prod_{e(w, w') \in \lambda} \hat{\pi}_e (w'\ |\ w) \\
	=\ &\prod_{e(w, w') \in \lambda} p_{\Lambda}(\Lambda(e(w, w'))\ |\ \Lambda(w)) 
\end{align*}
\noindent{where, without ambiguity, we let $p_{\Lambda}(..)$ denote the probability of an event in~${\cal{C}}_{\Lambda}$. Then}
\begin{align*}
\hat{\pi}_e (w'\ |\ w) &= p_{\Lambda}(\Lambda(e(w,w'))\ |\ \Lambda(w)) \\
	&= p(\Lambda(e(w, w'))\ |\ \Lambda(w), \Lambda) \\
	&= \frac{p(\Lambda(e(w, w')), \Lambda(w), \Lambda)}{p(\Lambda(w), \Lambda)} \\
&= \frac{p(\Lambda\ |\ \Lambda(e(w, w')), \Lambda(w))\ p(\Lambda(e(w, w')), \Lambda(w))}{p(\Lambda\ |\ \Lambda(w))\ p(\Lambda(w))} \\
&= \frac{p(\Lambda\ |\ \Lambda(e(w, w')))\ p(\Lambda(e(w, w')), \Lambda(w))}{p(\Lambda\ |\ \Lambda(w))\ p(\Lambda(w))} \\
\intertext{since $\Lambda(e(w, w')) \subset \Lambda(w)$}
&= \frac{p(\Lambda\ |\ \Lambda(e(w, w')))}{p(\Lambda\ |\ \Lambda(w))}\ \pi_e(w'\ |\ w)
\end{align*}

\noindent{Under this edge-probability assignment, no edges in ${\cal{C}}_{\Lambda}$ are given a zero probability, since each $e(w,w') \in \lambda \in \Lambda$.
And no position in ${\cal{C}}_{\Lambda}$ needs to be split (uncoalesced) in order for us to make this edge-probability assignment.}

By construction two vertices in a tree on the same route cannot be in the same position. So consider two vertices in ${\cal{C}}_{\Lambda}$ which do not lie
	on the same route. Then the collections of routes (elements of $\Lambda$) passing through each of these vertices are disjoint. So we can assign the
	probability distribution over these routes (in $\cal{C}$) in such a way that the conditional joint probability distributions on the subpaths
	emanating from these two vertices in ${\cal{C}}_{\Lambda}$ are different. Hence our assignment does not require us to coalesce distinct 
	positions in ${\cal{C}}_{\Lambda}$.

So position-structure is preserved.

Hence ${\cal{C}}_{\Lambda}$ is an sCEG, and the set of sCEGs is closed under conditioning on an intrinsic event.

\bigskip
\noindent{\bf Proof of Lemma~1:}

\smallskip
\noindent{Variables $X, Y$ partition the set of atoms of $\cal{C}$, and since $\Lambda \subset \Lambda(C)$, $X, Y$ also partition the set of 
atoms of ${\cal{C}}_{\Lambda}$.} 

\noindent{Consider arbitrary events $\Lambda_x, \Lambda_y$ from $\{ \Lambda_x \}_{x \in \mathbb{X}},\ \{ \Lambda_y \}_{y \in \mathbb{Y}}$, and the event
$\Lambda_x \cap \Lambda_y$. Then $p (\Lambda_x\ |\ \Lambda) = p_{\Lambda}(\Lambda_x)$ etc., and the statement}
\begin{align*}
p (\Lambda_x, \Lambda_y\ |\ \Lambda) &= p (\Lambda_x\ |\ \Lambda)\ p (\Lambda_y\ |\ \Lambda) \\
\intertext{is true if and only if the statement}
p_{\Lambda} (\Lambda_x, \Lambda_y) &= p_{\Lambda} (\Lambda_x)\ p_{\Lambda} (\Lambda_y) 
\end{align*}
\noindent{is true.
If either of these relationships holds for all $\Lambda_x \in \{ \Lambda_x \}_{x \in \mathbb{X}}$, $\Lambda_y \in \{ \Lambda_y \}_{y \in \mathbb{Y}}$,
then so does the other for all $\Lambda_x \in \{ \Lambda_x \}_{x \in \mathbb{X}}$, $\Lambda_y \in \{ \Lambda_y \}_{y \in \mathbb{Y}}$.}

\noindent{Hence $X \amalg Y\ |\ \Lambda$ if and only if $X \amalg Y$ in ${\cal{C}}_{\Lambda}$.}\hfill$\square$}

\bigskip
\noindent{\bf Proof of Lemma~2:}

\smallskip
\noindent{{\bf 1.} Consider a single route $\lambda$ consisting of a subpath $\mu_0(w_0, w)$ between $w_0$~and~$w$, the edge $e(w, w')$ labelled $x\ (\ne 0)$, and
a subpath $\mu_1(w', w_{\infty})$ connecting $w'$~to~$w_{\infty}$. Now this route consists of a set of edges and by construction the probability 
$p(\lambda)$ of the route is equal to the product of the probabilities labelling each of these edges. Moreover, the probability of any subpath
of $\lambda$ is equal to the product of the probabilities labelling each of its edges. So $p(\lambda)$ can be written as the product of the
probabilities of three subpaths: $\mu_0(w_0, w), e(w, w')$ and $\mu_1(w', w_{\infty})$. Thus:}
$$p(\lambda) = \pi_{\mu_0}(w\ |\ w_0)\ \pi_e (w'\ |\ w)\ \pi_{\mu_1}(w_{\infty}\ |\ w').$$
But the fact that $\lambda$ utilises the subpath $\mu_0(w_0, w)$ between $w_0$ and $w$ allows us to completely specify the value of the vector  
$\boldsymbol{X}_{U(w)}$. By a slight abuse of notation we can represent this as $\boldsymbol{X}_{U(w)} = \mu_0(w_0, w)$.

Consider now the event $(\boldsymbol{X}_{U(w)} = \mu_0(w_0, w), I(w) = 1, X(w) = x)$, which is the union of all $w_0 \rightarrow w_{\infty}$
routes which utilise the subpath $\mu_0(w_0, w)$ and the edge $e(w, w')$. Then since this is an intrinsic event we can write:
\begin{align*}
p(\boldsymbol{X}_{U(w)} = \mu_0(w_0, w),\ &I(w) = 1, X(w) = x) \\
	&= p(\Lambda(\mu_0(w_0,w)), \Lambda(w), \Lambda(e(w, w'))) \\
	&= \pi_{\mu_0}(w\ |\ w_0)\ \pi_e (w'\ |\ w) \sum_{\mu_1 \in M_1} \pi_{\mu_1}(w_{\infty}\ |\ w'),
\end{align*}
where $M_1$ is the set of all subpaths from $w'$ to $w_{\infty}$. But $\sum_{\mu_1 \in M_1} \pi_{\mu_1}(w_{\infty}\ |\ w')~=~1$ 
since all paths through $w'$ terminate in $w_{\infty}$. 

Similarly, for the event $(\boldsymbol{X}_{U(w)} = \mu_0(w_0, w), I(w) = 1)$ we have:
$$p(\boldsymbol{X}_{U(w)} = \mu_0(w_0, w), I(w) = 1) = p(\Lambda(\mu_0(w_0,w)), \Lambda(w)) = \pi_{\mu_0}(w\ |\ w_0).$$
So
\begin{align*}
p(X(w) = x\ |\ &\boldsymbol{X}_{U(w)} = \mu_0(w_0, w), I(w) = 1) = \frac{\pi_{\mu_0}(w\ |\ w_0)\ \pi_e(w'\ |\ w)}{\pi_{\mu_0}(w\ |\ w_0)} \\
	&= \pi_e(w'\ |\ w) = p(\Lambda(e(w, w'))\ |\ \Lambda(w)) \\
	&= p(X(w) = x\ |\ I(w) = 1).
\end{align*}
Hence 
\begin{align*}
&X(w) \amalg \boldsymbol{X}_{U(w)} \ |\ (I(w) = 1) &(1)
\end{align*}

\noindent{{\bf 2.} If $I(w) = 1$ then $X(w') = 0$ for all $w' \in D^c(w) \cap U^c(w)$, so we can completely specify the value of the vector
$\boldsymbol{X}_{D^c(w) \cap U^c(w)}$, and expression (1) implies:}
\begin{align*}
&X(w) \amalg \boldsymbol{X}_{U(w)} \ |\ (\boldsymbol{X}_{D^c(w) \cap U^c(w)}, I(w) = 1) &(2)
\end{align*}

\noindent{Moreover if $I(w) = 1$, no further information about $\boldsymbol{X}_{D^c(w) \cap U^c(w)}$ will assist us in predicting the value of $X(w)$.
Hence}
\begin{align*}
&X(w) \amalg \boldsymbol{X}_{D^c(w) \cap U^c(w)}\ |\ (I(w) = 1)  &(3)
\end{align*}

\noindent{Using a result from~\cite{Dawid1979}, the expressions (2) and (3) yield the result:}
\begin{align*} 
&X(w) \amalg (\boldsymbol{X}_{U(w)}, \boldsymbol{X}_{D^c(w) \cap U^c(w)}\ |\ (I(w) = 1) \\ 
\Rightarrow\ &X(w) \amalg \boldsymbol{X}_{D^c(w)}\ |\ (I(w) = 1)
\end{align*}
\noindent{{\bf 3.} If $I(w) = 0$ then $X(w) = 0$, and no further information about $\boldsymbol{X}_{D^c(w)}$ will assist us in predicting the
value of $X(w)$. Hence also}
\begin{align*} 
&X(w) \amalg \boldsymbol{X}_{D^c(w)}\ |\ (I(w) = 0)	&\square
\end{align*}

\noindent{\bf Proof of Theorem~2:}

\smallskip
\noindent{{\bf 1. Sufficient conditions for independence.} The sufficient conditions for independence are an almost immediate consequence of
similar results for Markov processes, but we include a proof here for completeness.}

\bigskip
\noindent{Consider an sCEG $\cal{C}$, and two positions $w_1, w_2 \in V(C) \backslash \{w_{\infty}\}$ such that\break 
$w_1 \prec w \prec w_2$ for some cut-vertex $w$. By construction $I(w_1) \not\equiv 0, I(w_2) \not\equiv 0,\break I(w) \equiv 1$.}

\noindent{Consider the event $(X(w_1) = x_1, X(w_2) = x_2) \equiv (X(w_1) = x_1, I(w) = 1,\break X(w_2) = x_2)$ for $x_1 \ne 0, x_2 \ne 0$.
This is the union of all routes passing through~$w_1$, utilising an edge $e(w_1, w_1')$ labelled~$x_1$, passing though~$w$, passing
through~$w_2$, and utilising an edge $e(w_2, w_2')$ labelled~$x_2$. By analogy with the proof of Lemma~2 we can, since this is an
intrinsic event, write:}
\begin{align*}
p(X(w_1&) = x_1, X(w_2) = x_2) = \sum_{\mu_0 \in M_0} \pi_{\mu_0} (w_1\ |\ w_0)\ \pi_e(w_1'\ |\ w_1) 
		\sum_{\mu_1 \in M_1} \pi_{\mu_1} (w\ |\ w_1') \\
	&\times \sum_{\mu_2 \in M_2} \pi_{\mu_2} (w_2\ |\ w)\ \pi_e(w_2'\ |\ w_2) \sum_{\mu_3 \in M_3} \pi_{\mu_3} (w_{\infty}\ |\ w_2') \\
\intertext{where $M_0$ is the set of all subpaths from $w_0$ to $w_1$, $M_1$ is the set of all subpaths from $w_1'$ to $w$,
$M_2$ is the set of all subpaths from $w$ to $w_2$, and $M_3$ is the set of all subpaths from $w_2'$ to $w_{\infty}$. 
But $\sum_{\mu_0 \in M_0} \pi_{\mu_0} (w_1\ |\ w_0)$ is simply the probability of reaching $w_1$ from $w_0$ etc., so
this equals} 
	&= \pi (w_1\ |\ w_0)\ \pi_e(w_1'\ | w_1)\ \pi (w\ |\ w_1')\ \pi (w_2\ |\ w)\ \pi_e(w_2'\ |\ w_2)\ \pi (w_{\infty}\ |\ w_2') \\ 
	&= \pi (w_1\ |\ w_0)\ \pi_e(w_1'\ | w_1)\times 1 \times \pi (w_2\ |\ w)\ \pi_e (w_2'\ |\ w_2) \times 1 
\end{align*}
\noindent{Similarly for the event $X(w_1) = x_1$, we can write:
\begin{align*}
p(X(w_1) = x_1) &= \pi(w_1\ |\ w_0)\ \pi_e(w_1'\ |\ w_1) \times 1 \\
\intertext{so}
p(X(w_2) = x_2\ |\ X(w_1) = x_1) &= \frac{\pi (w_1\ |\ w_0)\ \pi_e(w_1'\ | w_1)\ \pi (w_2\ |\ w)\ \pi_e (w_2'\ |\ w_2)}
		{\pi (w_1\ |\ w_0)\ \pi_e(w_1'\ | w_1)} \\
	&= \pi (w_2\ |\ w)\ \pi_e (w_2'\ |\ w_2)
\end{align*}
\noindent{Now consider the event $(X(w_2) = x_2) \equiv (I(w) = 1, X(w_2) = x_2)$. Analogously with above we can write:}
\begin{align*}
p(X(w_2) = x_2) &= \pi (w\ |\ w_0)\ \pi (w_2\ |\ w)\ \pi_e(w_2'\ |\ w_2)\ \pi (w_{\infty}\ |\ w_2') \\ 
	&= 1 \times \pi (w_2\ |\ w)\ \pi_e (w_2'\ |\ w_2) \times 1 \\
	&= p(X(w_2) = x_2\ |\ X(w_1) = x_1)
\end{align*}

\noindent{It is straightforward to show that the result also holds for $x_1 = 0$ and $x_2 = 0$.}


\smallskip
\noindent{If $w_2$ is itself a cut-vertex (with $w_1 \prec w_2$), then we replace $I(w) = 1$ by $I(w_2) = 1$ in the above argument with the same result.}

So a sufficient condition for $X(w_1) \amalg X(w_2)$ is that either $w_2$ is itself a cut-vertex, or there exists a cut-vertex~$w$ such
that $w_1 \prec w \prec w_2$.

\bigskip
\noindent{\bf 2. Necessary conditions for independence.} 

\smallskip
Let $X(w_1) \amalg X(w_2)$ (and since $I(w)$ is a function of $X(w)$, $X(w_1) \amalg I(w_2)$ and 
$I(w_1) \amalg I(w_2)$). Let the set of routes of $\cal{C}$ 
be partitioned into four subsets. Call a route Type $A$ if it passes through $w_2$, but not through $w_1$,
Type $B$ if it passes through neither $w_1$ nor~$w_2$,
Type $C$ if it passes through both $w_1$ and $w_2$, and
Type $D$ if it passes through~$w_1$, but not through $w_2$.
Our proof proceeds as follows:

\noindent{(a) We show that we must have $w_1 \prec w_2$ (ie. the set of Type~$C$ routes is non-empty).}

\noindent{(b) We show that every route intersects with every other route at some point downstream of $w_0$ and upstream of $w_{\infty}$.
If two $w_0 \rightarrow w_{\infty}$ routes share no vertices except $w_0$ and $w_{\infty}$, we call them
{\it internally disjoint}. So there cannot be two internally disjoint $w_0 \rightarrow w_{\infty}$ routes in $\cal{C}$}

\noindent{(c) We show that there must therefore be a cut-vertex between $w_0$ and $w_{\infty}$.}

\noindent{(d) We show that either $w_1$ is a cut-vertex or $w_2$ is a cut-vertex, or there exists a cut-vertex $w$ such that 
$w_1 \prec w \prec w_2$.}

\noindent{(e) Finally we show that if $w_1$ is a cut-vertex then there must also either be a cut-vertex at $w_2$ or a cut-vertex $w$ such that 
$w_1 \prec w \prec w_2$.}

\smallskip
(a) Suppose that $w_1 \not\prec w_2$ (and recall that $w_2 \not\prec w_1$). Then\break 
$p(I(w_2) = 1\ |\ I(w_1) = 1) \equiv 0$. 
$I(w_1) \amalg I(w_2) \Rightarrow p(I(w_2) = 1) \equiv 0\break \Rightarrow I(w_2) \equiv 0$. This is impossible by
construction. Therefore $w_1 \prec w_2$.  

(b) We first show that each Type $C$ route intersects with every other route at $w_1$ or at $w_2$
or at some point between these positions.

\noindent{Let $\lambda_1$ be a Type $C$ route, and $\mu_1(w_1, w_2)$ the subpath coincident with $\lambda_1$ between
$w_1$ and $w_2$. If the set of Type~$B$ routes is non-empty then let $\lambda_2$ be a Type~$B$ route which does not intersect
with $\mu_1$ (ie. $\lambda_2$ and $\mu_1$ have no positions or edges in common).}

\noindent{Consider a distribution $P$ which assigns (1)~a probability of $1 - \epsilon$ to every edge of the subpath $\mu_1(w_1, w_2)$, 
and (2)~a probability of $1 - \delta$ to each edge of the route~$\lambda_2$. Let the number of edges in $\mu_1(w_1, w_2)$ be $n(\mu_1)$
and the number of edges in $\lambda_2$ be $n(\lambda_2)$ (where both $n(\mu_1)$ and $n(\lambda_2)$ are finite). Then let
$(1 - \epsilon)^{n(\mu_1)} > 0.9$ and $(1 - \delta)^{n(\lambda_2)} > 0.8$. If $\lambda_2$ does not intersect with $\mu_1$ then this is always 
possible.}

\smallskip
Under $P$, assignment~(1) gives us that
\begin{align*}
&p(I(w_2) = 1\ |\ I(w_1) = 1) \ge (1 - \epsilon)^{n(\mu_1)} > 0.9 \\
\intertext{and $I(w_1) \amalg I(w_2)$ implies that under this $P$}
&p(I(w_2) = 1\ |\ I(w_1) = 0) > 0.9 \ \Rightarrow\ p(I(w_2) = 0\ |\ I(w_1) = 0) < 0.1 \\
\intertext{But assignment (2) gives us that}                
&p(I(w_2) = 0) \ge p(I(w_1) = 0, I(w_2) = 0) \ge p(\lambda_2) = (1 - \delta)^{n(\lambda_2)} > 0.8 \\
\Rightarrow\ &p(I(w_2) = 0\ |\ I(w_1) = 0) > 0.8 \quad\quad\quad \hbox{\textreferencemark}
\end{align*}

\noindent{The assumption $I(w_1) \amalg I(w_2)$ is incompatible with the assignments of (1) and~(2). But these assignments
are always possible if $\lambda_2$ does not intersect with $\mu_1$. Hence $\lambda_2$ must intersect with $\mu_1$.}

\noindent{Hence each Type $C$ route intersects with every Type $B$ route at some point downstream of
$w_1$ and upstream of $w_2$.
Also each Type~$C$ route intersects with every Type~$A$ route (at $w_2$), with every Type~$D$ route (at $w_1$)
and with every other Type~$C$ route (at both $w_1$ and $w_2$).}

We now consider routes that are not of Type~$C$.
If the set of non-Type~$C$ routes is non-empty let $\lambda_3, \lambda_4$ be members of this set which do not 
intersect except at $w_0$ and $w_{\infty}$. Let $\mu(w_1, w_2)$ be a subpath between $w_1$ and $w_2$.

\noindent{From above both $\lambda_3$ and $\lambda_4$ must intersect with $\mu$. Let $\lambda_3$ intersect with~$\mu$ only
at the positions $w_{31}, \dots w_{3m}$, where $w_{31} \prec \dots \prec w_{3m}$; and let
$\lambda_4$ intersect with $\mu$ only
at the positions $w_{41}, \dots w_{4n}$, where $w_{41} \prec \dots \prec w_{4n}$. Without
loss of generality let $w_1 \preceq w_{31} \prec w_{41} \preceq w_2$, so that $\lambda_3$ could be a route of Type~$B$ or
Type~$D$, and $\lambda_4$ could be a route of Type~$A$ or Type~$B$.}

\noindent{Suppose firstly that $w_{4n} \prec w_{3m}$. Consider the subpath $\mu_5(w_1, w_2)$ which coincides with $\mu$ from 
$w_1$ to $w_{31}$ (if $w_{31} \ne w_1$), coincides with $\lambda_3$ from $w_{31}$ to $w_{3m}$, and
coincides with $\mu$ from $w_{3m}$ to $w_2$. 
This subpath $\mu_5$ does not intersect with the route~$\lambda_4$. This is impossible since every route
in~$\cal{C}$ intersects with every $\mu(w_1, w_2)$ subpath.}

\noindent{Suppose therefore that $w_{3m} \prec w_{4n}$. Consider the subpath $\mu_6(w_1, w_{\infty})$ which
coincides with $\mu$ from $w_1$ to $w_{31}$ (if $w_{31} \ne w_1$) and coincides with $\lambda_3$ from $w_{31}$ to 
$w_{\infty}$; and the subpath $\mu_7(w_0, w_2)$ which coincides with $\lambda_4$ from $w_0$ to $w_{4n}$ and
coincides with $\mu$ from $w_{4n}$ to $w_2$ (if $w_{4n} \ne w_2$).
Consider also a distribution $P$ which assigns (1)~a probability of $1 - \epsilon$ to every edge of~$\mu_6$, 
and (2)~a probability of $1 - \delta$ to every edge of $\mu_7$. Let the number of edges in $\mu_6(w_1, w_{\infty})$ be $n(\mu_6)$
and the number of edges in $\mu_7(w_0, w_2)$ be $n(\mu_7)$ (where both $n(\mu_6)$ and $n(\mu_7)$ are finite). Then let
$(1 - \epsilon)^{n(\mu_6)} > 0.9$ and $(1 - \delta)^{n(\mu_7)} > 0.8$. If $\lambda_3$ and $\lambda_4$ do not intersect then this is always 
possible.}

\smallskip
Under $P$, assignment~(1) gives us that
\begin{align*}
&p(I(w_2) = 0\ |\ I(w_1) = 1) \ge (1 - \epsilon)^{n(\mu_6)} > 0.9 \\
\intertext{and $I(w_1) \amalg I(w_2)$ implies that under this $P$}
&p(I(w_2) = 0\ |\ I(w_1) = 0) > 0.9 \ \Rightarrow\ p(I(w_2) = 1\ |\ I(w_1) = 0) < 0.1 \\
\intertext{But assignment (2) gives us that}                
&p(I(w_2) = 1\ |\ I(w_1) = 0) > 0.8 \quad\quad\quad \hbox{\textreferencemark}
\end{align*}

\noindent{The assumption $I(w_1) \amalg I(w_2)$ is incompatible with the assignments of (1) and~(2). But these assignments
are always possible if $\lambda_3$ and $\lambda_4$ do not intersect. Hence $\lambda_3$ and $\lambda_4$ must intersect.}

\noindent{Hence each Type $B$ route intersects with every Type $A$, Type~$B$ or Type~$D$ route, and each Type~$A$ route
intersects with every Type~$D$ route.
Also, each Type~$A$ route intersects with every other Type~$A$ route (at $w_2$), and each Type~$D$ route intersects
with every other Type~$D$ route (at $w_1$). So each route in $\cal{C}$ intersects with every other route downstream of
$w_0$ and upstream of $w_{\infty}$.}

Hence there cannot be two internally disjoint directed routes from $w_0$ to~$w_{\infty}$. 

(c) To show that this implies the existence of a cut-vertex between $w_0$ and~$w_{\infty}$, we briefly consider a CEG as a Flow Network where
every edge and every vertex (except $w_0$ and $w_{\infty}$) has a (flow) capacity of one. Then the maximum flow through the CEG from 
$w_0$ to $w_{\infty}$ must equal the maximum number of internally disjoint $w_0 \rightarrow w_{\infty}$ routes.
We can now use Ford \& Fulkersons' Max Flow Min Cut Theorem~\cite{FandF}. This theorem applies to networks where only the edges are given
capacities, so we replace each vertex $w \in V({\cal{C}}) \backslash \{ w_0, w_{\infty} \}$ by a pair of vertices $w^-, w^+$ connected by an
edge $e(w^-, w^+)$ with a capacity of one -- the only edge emanating from $w^-$ being $e(w^-, w^+)$ and the only edge entering $w^+$ being
$e(w^-, w^+)$.

The theorem states that for a Flow Network with a single source and a single sink, the maximum flow from source to sink equals the capacity
of the minimum cut, where cuts pass through the {\bf edges} of the graph (ie. a cut partitions $V({\cal{C}})$ into two collections of vertices
with $w_0$ in one collection and $w_{\infty}$ in the other), and the capacity of the minimum cut is the sum of the capacities of the edges which 
are cut.

So if in our CEG we have no pairwise internally disjoint $w_0 \rightarrow w_{\infty}$ routes, then the maximum flow through the CEG from $w_0$
to $w_{\infty}$ must equal one, and the capacity of the minimum cut of the CEG must also equal one.
Hence all $w_0 \rightarrow w_{\infty}$ routes must pass through a single edge. 

Now this edge may be of the form $e(w^-, w^+)$, in which case $w$ is a cut-vertex; or the edge may be of the form $e(w_a, w_b)$ for $w_a \ne w_b$,
in which case both $w_a$ and $w_b$ are cut-vertices. Hence there is a cut-vertex $w$ such that $w_0 \prec w \prec w_{\infty}$.

\smallskip
\noindent{This result can also be arrived at by using a corollary of Whitney's~\cite{Whitney} Theorem~7 (a result for undirected
graphs, sometimes described as the {\it 2nd variation of Menger's Theorem}~\cite{Menger}).}

(d) Suppose there exists a cut-vertex upstream of $w_1$. Then relabel this cut-vertex as $w_0$ and repeat the
argument of (b)(c) to show that there exists a cut-vertex between this new $w_0$ and $w_{\infty}$. Since the number of 
positions in $\cal{C}$
is finite, repeated use of this argument shows us that either $w_1$ is a cut-vertex or there exists a cut-vertex downstream of~$w_1$. 
A complementary argument shows that there exists a cut-vertex at $w_2$ or upstream of $w_2$.

(e) Suppose $w_1$ is a cut-vertex, but $w_2$ is not. Then either (i) $w_2$ lies exactly one edge downstream of $w_1$ on every $w_1 \rightarrow w_2$
subpath, or (ii) there exists a position $w^1_1$ ($\ne w_2$) exactly one
edge downstream of $w_1$ lying on a $w_1 \rightarrow w_2$ subpath. 

\noindent{(i) We know that $X(w_1) \ne 0$ (since $w_1$ is a cut-vertex), so if $X(w_1)$ takes a value corresponding to an edge from $w_1$ to $w_2$,
then $I(w_2) = 1$ and $X(w_2) > 0$; otherwise $I(w_2) = X(w_2) = 0$. So $X(w_2) \; {\hbox{{\bf /}\llap{$\amalg$}}}\; X(w_1)$. $\quad$
\textreferencemark}

\noindent{(ii) If $X(w_1)$ takes a value corresponding to an edge from $w_1$ to $w^1_1$, then\break $I(w^1_1) = 1$; otherwise $I(w^1_1) = 0$. Hence
$I(w^1_1)$ is a function of $X(w_1)$. So $X(w_1) \amalg X(w_2) \Rightarrow X(w_1) \amalg I(w_2) \Rightarrow I(w^1_1) \amalg I(w_2)$, and using
the argument of (b), (c), (d) above there must be a cut-vertex at $w^1_1$ or between $w^1_1$ and $w_2$.}

Therefore there exists a cut-vertex at $w_2$ or a cut-vertex $w$ such that\break $w_1 \prec w \prec w_2$. 

\hfill$\square$

\bigskip
\noindent{\bf Proof of Corollary~1:}

\smallskip
\noindent{Let $X(w_1) \amalg X(w_2)$ hold for some $w_1 \in W_a,\ w_2 \in W_b$. Then by Theorem~2 either (i)~$w_2$ is a cut-vertex (in which case
$W_b$ consists of the one position~$w_2$), or (ii)~there exists a cut-vertex $w$ such that $w_1 \prec w \prec w_2$.

Since $W_a$ and $W_b$ are position cuts, this implies that either (i)~$w_a \prec w_2\break \forall w_a \in W_a$, or (ii)~$w_a \prec w \prec w_b \;
\forall w_a \in W_a, w_b \in W_b$, and hence\break (i)~$X(w_a) \amalg X(w_2)  \; \forall w_a \in W_a$, or (ii)~$X(w_a) \amalg X(w_b)  \; \forall w_a \in W_a,
w_b \in W_b$.

\smallskip
Note that $X(w_a)$, $X(w_b)$ pairwise independent for all $w_a, w_b$ does not in general imply groupwise independence, but it does here:

\noindent{Any event characterised by the expression $\boldsymbol{X}_{W_a} = \boldsymbol{x}_a$ has the form:
$$X(w_a') = x_a\; (\ne 0)\ \hbox{for some}\ w_a' \in W_a,\quad X(w_a) = 0\quad \forall w_a \in W_a \backslash \{w_a' \}$$
So}
\begin{align*}
p(\boldsymbol{X}_{W_a} =\ &\boldsymbol{x}_a, \boldsymbol{X}_{W_b} = \boldsymbol{x}_b) \\
	=\ &p(X(w_a') = x_a, X(w_b') = x_b, X(w) = 0\quad \forall w \in W_a \cup W_b \backslash \{w_a', w_b' \}) \\
\intertext{for some $w_a' \in W_a, w_b' \in W_b$}
	=\ &p(X(w_a') = x_a, X(w_b') = x_b) \\
\intertext{since $X(w_a') \ne 0 \Rightarrow X(w_a) = 0\quad \forall w_a \in W_a \backslash \{w_a' \}$ etc}
 	=\ &p(X(w_a') = x_a)\ p(X(w_b') = x_b) \\
\intertext{since $X(w_a') \amalg X(w_b')$}
 	=\ &p(X(w_a') = x_a, X(w_a) = 0\quad \forall w_a \in W_a \backslash \{w_a' \})\\  
	&\times p(X(w_b') = x_b, X(w_b) = 0\quad \forall w_b \in W_b \backslash \{w_b' \}) \\
	=\ &p(\boldsymbol{X}_{W_a} = \boldsymbol{x}_a)\ p(\boldsymbol{X}_{W_b} = \boldsymbol{x}_b)
\end{align*}
\noindent{So $\boldsymbol{X}_{W_a} \amalg \boldsymbol{X}_{W_b}$.
But $X(W_a) = \sup_{w \in W_a} X(w_a)$ is a function of $\boldsymbol{X}_{W_a}$, and $X(W_b)$ is a function of 
$\boldsymbol{X}_{W_b}$. Hence $X(W_a) \amalg X(W_b)$.}

\hfill{$\square$}


\bigskip
\noindent{\bf Proof of Corollary~2:}

\smallskip
\noindent{Since $\Lambda$ is intrinsic to $\cal{C}$, ${\cal{C}}_{\Lambda}$ is a subgraph of $\cal{C}$ with $V({\cal{C}}_{\Lambda}) \subset V(\cal{C})$. 

\smallskip
\noindent{Let~$W_a$ in~${\cal{C}}_{\Lambda}$ be the subset of $V({\cal{C}}_{\Lambda})$ which consists of elements of~$W_a$ in~$\cal{C}$. Then $W_a$ is
well-defined on ${\cal{C}}_{\Lambda}$, as is $X(w_a)$ for any $w_a \in W_a$.} 

\smallskip
\noindent{$X(W_a)$ is measurable with respect to the sigma-field of $\cal{C}$, so it partitions the set of atoms of~$\cal{C}$. Since 
$\Lambda \subset \Lambda ({\cal{C}})$, it also partitions the set of atoms of~${\cal{C}}_{\Lambda}$, and is well-defined on~${\cal{C}}_{\Lambda}$ as:}
$$X(W_a)\ = \sup_{\hbox{\scriptsize
				$\begin{array}{c}
				w_a \in W_a \\ 
				w_a \in V({\cal{C}}_{\Lambda})
				\end{array}
					$}} X(w_a).
	$$
Hence $p_{\Lambda} (X(W_a) = x_a) = p(X(W_a) = x_a\ |\ \Lambda)$, and all necessary terms are defined on ${\cal{C}}_{\Lambda}$ consistently with 
their definitions on $\cal{C}$.

\smallskip
\noindent{In ${\cal{C}}_{\Lambda}$ there exists a cut-vertex $w$ such that $W_a \prec w \prec W_b$, so by Theorem~2, $X(w_a) \amalg X(w_b)$ holds
in~${\cal{C}}_{\Lambda}$ for any $w_a \in W_a \cap V({\cal{C}}_{\Lambda}), w_b \in W_b \cap V({\cal{C}}_{\Lambda})$.}

\smallskip
\noindent{Hence by Lemma~1, $X(W_a) \amalg X(W_b)\ |\ \Lambda$ holds in $\cal{C}$.} 

\hfill $\square$

     \bibliographystyle{plain}   
     \bibliography{newsepbib}   

\begin{thebibliography}{10}

\bibitem{Allmanrhodes}
E.~S. Allman, C.~Matias, and J.~A. Rhodes.
\newblock Identifiability of parameters in latent structure models with many
  observed variables.
\newblock {\em The Annals of Statistics}, 37:3099--3132, 2009.

\bibitem{AMP2}
S.~A. Andersson, D.~Madigan, and M.~D. Perlman.
\newblock Alternative {M}arkov properties for chain graphs.
\newblock {\em Scandinavian Journal of Statistics}, 28:33--85, 2001.

\bibitem{Lorna2}
L.~M. Barclay, J.~L. Hutton, and J.~Q. Smith.
\newblock Refining a {B}ayesian {N}etwork using a {C}hain {E}vent {G}raph.
\newblock {\em International Journal of Approximate Reasoning}, 54:1300--1309,
  2013.

\bibitem{Lorna1}
L.~M. Barclay, J.~L. Hutton, and J.~Q. Smith.
\newblock Chain {E}vent {G}raphs for {I}nformed {M}issingness.
\newblock {\em Bayesian Analysis}, 9:53--76, 2014.

\bibitem{Lornaetal}
L.~M. Barclay, J.~Q. Smith, P.~A. Thwaites, and A.~E. Nicholson.
\newblock The {D}ynamic {C}hain {E}vent {G}raph.
\newblock Research {R}eport 14-04, CRiSM, 2014.
\newblock Submitted to {\it {E}lectronic {J}ournal of {S}tatistics}.

\bibitem{BandMilan}
R.~R. Bouckaert and M.~Studeny.
\newblock Chain graphs: Semantics and expressiveness.
\newblock In C.~Froidevaux and J.~Kohlas, editors, {\em Symbolic and Quantative
  approaches to Reasoning and Uncertainty}, number 946 in Lecture Notes in
  Artificial Intelligence, pages 67--76. Springer-Verlag, 1995.

\bibitem{kollerUAI1996}
C.~Boutilier, N.~Friedman, M.~Goldszmidt, and D.~Koller.
\newblock Context-specific independence in {B}ayesian {N}etworks.
\newblock In {\em Proceedings of the 12th Conference on Uncertainty in
  Artificial Intelligence}, pages 115--123, 1996.

\bibitem{FirstPDGs}
M.~Bozga and O.~Maler.
\newblock On the {R}epresentation of {P}robabilities over {S}tructured
  {D}omains.
\newblock In {\em Computer Aided Verification}, volume 1633 of {\em Lecture
  Notes in Computer Science}, pages 261--273. Springer, 1999.

\bibitem{Cowelletal}
R.~G. Cowell, A.~P. Dawid, S.~L. Lauritzen, and D.~J. Spiegelhalter.
\newblock {\em Probabilistic Networks and Expert Systems}.
\newblock Springer, 1999.

\bibitem{RobandJim}
R.~G. Cowell and J.~Q. Smith.
\newblock Causal {D}iscovery through {MAP} selection of stratified {C}hain
  {E}vent {G}raphs.
\newblock {\em Electronic Journal of Statistics}, 8:965--997, 2014.

\bibitem{Dawid1979}
A.~P. Dawid.
\newblock Conditional independence in statistical theory.
\newblock {\em Journal of the Royal Statistical Society, Series B}, 41:1--31,
  1979.

\bibitem{DawidMilan}
A.~P. Dawid and M.~Studeny.
\newblock Conditional products: an alternative approach to conditional
  independence.
\newblock In {\em Proceedings of the 7th Workshop on Artificial Intelligence
  and Statistics}, pages 32--40, 1999.

\bibitem{FandF}
L.~R. Ford and D.~R. Fulkerson.
\newblock {\em Flows in Networks}.
\newblock Princeton University Press, 1962.

\bibitem{Guy}
G.~Freeman and J.~Q. Smith.
\newblock Bayesian {MAP} model selection of {C}hain {E}vent {G}raphs.
\newblock {\em Journal of Multivariate Analysis}, 102:1152--1165, 2011.

\bibitem{Guy2}
G.~Freeman and J.~Q. Smith.
\newblock Dynamic staged trees for discrete multivariate {T}ime {S}eries:
  {F}orecasting, model selection and causal analysis.
\newblock {\em Bayesian Analysis}, 6:279--305, 2011.

\bibitem{HandL}
S.~Hojsgaard and S.~L. Lauritzen.
\newblock Graphical {G}aussian models with edge and vertex symmetries.
\newblock {\em Journal of the Royal Statistical Society, Series B},
  70:1005--1027, 2008.

\bibitem{Jaeger04}
M.~Jaeger, J.~D. Nielsen, and T.~Silander.
\newblock Learning {P}robabilistic {D}ecision {G}raphs.
\newblock In {\em Proceedings of the 2nd European Workshop on Probabilistic
  Graphical Models}, pages 113--120, Leiden, 2004.

\bibitem{kallenberg}
O.~Kallenberg.
\newblock {\em Foundations of {M}odern {P}robability}.
\newblock Springer, 1997.

\bibitem{KorbandN}
K.~B. Korb and A.~E. Nicholson.
\newblock {\em Bayesian Artificial Intelligence}.
\newblock Chapman \& Hall/CRC Press, 2004.

\bibitem{lauritzenbook}
S.~L. Lauritzen.
\newblock {\em Graphical Models}.
\newblock Oxford, 1996.

\bibitem{Lauritzen2001}
S.~L. Lauritzen.
\newblock Causal inference from graphical models.
\newblock In O.~E. Barndorff-Nielsen et~al., editors, {\em Complex Stochastic
  Systems}. Chapman and Hall, 2001.

\bibitem{Laurdsep}
S.~L. Lauritzen, A.~P. Dawid, B.~N. Larsen, and H.~G. Leimer.
\newblock Independence properties of directed {M}arkov fields.
\newblock {\em Networks}, 20:491--505, 1990.

\bibitem{McAllester}
D.~McAllester, M.~Collins, and F.~Periera.
\newblock Case factor diagrams for structured probabilistic modeling.
\newblock In {\em Proceedings of the 20th Conference on Uncertainty in
  Artificial Intelligence}, pages 382--391, 2004.

\bibitem{Meek}
C.~Meek.
\newblock Strong completeness and faithfulness in {B}ayesian {N}etworks.
\newblock In {\em Proceedings of the 11th Conference on Uncertainty in
  Artificial Intelligence}, pages 411--441, 1995.

\bibitem{Menger}
K.~Menger.
\newblock Zur allgemeinen {K}urventheorie.
\newblock {\em Fundamenta Mathematicae}, 10:95--115, 1927.

\bibitem{MondandJim}
D.~M.~Q. Mond, J.~Q. Smith, and D.~{Van Straten}.
\newblock Stochastic factorisations, sandwiched simplices and the topology of
  the space of explanations.
\newblock {\em Proceedings of the Royal Society of London}, A 459:2821--2845,
  2003.

\bibitem{Olmstead}
S.~M. Olmstead.
\newblock {\em On {R}epresenting and {S}olving {D}ecision {P}roblems}.
\newblock PhD thesis, Stanford University, 1983.

\bibitem{Pearl2000}
J.~Pearl.
\newblock {\em Causality: Models, Reasoning and Inference}.
\newblock Cambridge, 2000.

\bibitem{PooleandZ}
D.~Poole and N.~L. Zhang.
\newblock Exploiting contextual independence in probabilistic inference.
\newblock {\em Journal of Artificial Intelligence Research}, 18:263--313, 2003.

\bibitem{TomandSp}
T.~S. Richardson and P.~Spirtes.
\newblock Ancestral graph {M}arkov models.
\newblock {\em Annals of Statistics}, 30:962--1030, 2002.

\bibitem{SalCanM}
A.~Salmeron, A.~Cano, and S.~Moral.
\newblock Importance sampling in {B}ayesian {N}etworks using probability trees.
\newblock {\em Computational Statistics and Data Analysis}, 34:387--413, 2000.

\bibitem{Shafer}
G.~Shafer.
\newblock {\em The Art of Causal Conjecture}.
\newblock MIT Press, 1996.

\bibitem{SilanderCEG}
T.~Silander and T-Y. Leong.
\newblock A {D}ynamic {P}rogramming {A}lgorithm for {L}earning {C}hain {E}vent
  {G}raphs.
\newblock In {\em Discovery Science}, volume 8140 of {\em Lecture Notes in
  Computer Science}, pages 201--216. Springer, 2013.

\bibitem{PaulandJim}
J.~Q. Smith and P.~E. Anderson.
\newblock {C}onditional independence and {C}hain {E}vent {G}raphs.
\newblock {\em Artificial Intelligence}, 172:42--68, 2008.

\bibitem{JandPDTs}
J.~Q. Smith and P.~A. Thwaites.
\newblock Decision trees.
\newblock In E.~L. Melnick and B.~S. Everitt, editors, {\em Encyclopedia of
  Quantitative {R}isk Analysis and Assessment}, volume~2, pages 462--470.
  Wiley, 2008.

\bibitem{Spirtes}
P.~Spirtes, C.~Glymour, and R.~Scheines.
\newblock {\em Causation, Prediction and Search}.
\newblock Springer-Verlag, 1993.

\bibitem{newcap}
P.~A. Thwaites.
\newblock Causal identifiability via {C}hain {E}vent {G}raphs.
\newblock {\em Artificial Intelligence}, 195:291--315, 2013.

\bibitem{pgm}
P.~A. Thwaites and J.~Q. Smith.
\newblock Evaluating causal effects using {C}hain {E}vent {G}raphs.
\newblock In {\em Proceedings of the 3rd European Workshop on Probabilistic
  Graphical Models (PGM)}, pages 291--300, Prague, 2006.

\bibitem{ipmu}
P.~A. Thwaites and J.~Q. Smith.
\newblock Non-symmetric models, {C}hain {E}vent {G}raphs and {P}ropagation.
\newblock In {\em Proceedings of the 11th International Conference on
  Information Processing and Management of Uncertainty in Knowledge-Based
  Systems (IPMU)}, pages 2339--2347, Paris, 2006.

\bibitem{Crismsep}
P.~A. Thwaites and J.~Q. Smith.
\newblock Separation theorems for {C}hain {E}vent {G}raphs.
\newblock Research {R}eport 11-09, CRiSM, 2011.

\bibitem{uai2008}
P.~A. Thwaites, J.~Q. Smith, and R.~G. Cowell.
\newblock Propagation using {C}hain {E}vent {G}raphs.
\newblock In {\em Proceedings of the 24th Conference on Uncertainty in
  Artificial Intelligence}, pages 546--553, Helsinki, 2008.

\bibitem{CausalAI}
P.~A. Thwaites, J.~Q. Smith, and E.~M. Riccomagno.
\newblock Causal analysis with {C}hain {E}vent {G}raphs.
\newblock {\em Artificial Intelligence}, 174:889--909, 2010.

\bibitem{VandP}
T.~Verma and J.~Pearl.
\newblock Causal networks: semantics and expressiveness.
\newblock In {\em Proceedings of the 4th Conference on Uncertainty in
  Artificial Intelligence}, pages 352--359, 1988.

\bibitem{Whitney}
H.~Whitney.
\newblock Congruent {G}raphs and the {C}onnectivity of {G}raphs.
\newblock {\em American Journal of Mathematics}, 54(1):150--168, 1932.

\end{thebibliography}

\end{document}